\documentclass[fleqn,showpacs,showkeys,twocolumn]{revtex4}
\usepackage{graphicx}
\usepackage{amssymb}
\usepackage{amsmath}
\usepackage{bm}
\begin{document}

\title{Formulations of the Einstein equations for numerical simulations
\footnote{Invited Lecture at APCTP Winter School on Black Hole Astrophysics, 
Daejeon and Pohang, Korea, January 24-29, 2008.}}
\author{Hisa-aki Shinkai}
\email{shinkai@is.oit.ac.jp}
\affiliation{Department of Information Systems, 
Faculty of Information Science and Technology, 
Osaka Institute of Technology, 
Kitayama 1-79-1, Hirakata, Osaka 573-0196, Japan
}

\date[]{May 1, 2008}

\begin{abstract}
We review recent efforts to re-formulate the Einstein equations
for fully relativistic numerical simulations.  
The so-called numerical relativity 
is a promising research field matching with ongoing 
gravitational wave observations. 
In order to complete a long-term and accurate simulations of binary compact objects, 
people seek a robust set of equations against the violation of constraints. 
Many trials have revealed that mathematically 
equivalent sets of evolution equations show different numerical 
stability in free evolution schemes.  
In this article, we overview the efforts of the community,  
categorizing them into three directions: 
(1) modifications of the standard Arnowitt-Deser-Misner equations
initiated by the Kyoto group (the so-called Baumgarte-Shapiro-Shibata-Nakamura equations), 
(2) rewriting the evolution equations in a hyperbolic form, and 
(3) construction of an ``asymptotically constrained" system.  
We then introduce our series of works that tries to explain these evolution 
behaviors in a unified way using 
eigenvalue analysis of the constraint propagation equations.  
The modifications of (or adjustments to) the evolution 
equations change the character of constraint propagation, and 
several particular adjustments using constraints are expected to damp the 
constraint-violating modes. 
We show several set of adjusted ADM/BSSN equations, together with their numerical 
demonstrations. 
\end{abstract}



\maketitle

\section{Introduction}
\subsection{Overview}

The theory of general relativity describes the nature of the 
strong gravitational field. 
The Einstein equation predicts quite unexpected phenomena such as 
gravitational collapse, gravitational waves, the expanding universe and 
so on, which are all attractive not only for researchers but also for the public. 
The Einstein equation consists of 10 partial differential equations (elliptic
and hyperbolic) for 10 metric components, and it is not easy to solve 
them for any particular situation.  
Over the decades, people have tried to study the general-relativistic 
world by finding its
exact solutions, by developing approximation methods, or by simplifying the situations. 
While ``The Exact Solution" book \cite{exactsolution} says there were more than 
4000 publications on exact solutions between 1980 and 2000,   
direct numerical integration  of the Einstein equations
can be said to be the most robust way to study the strong gravitational field. 
This research field is often called ``numerical relativity". 

With the purpose of the predictions of precise gravitational waveforms
from coalescences of the binary neutron-stars and/or black-holes, 
numerical relativity has been developed for the past three decades. 
The difficulty of numerical integrations of the Einstein equations 
arises from its mathematical complexity of the equations, physical 
difficulty of singularity treatments,   
and from high-level requirements for computational skills and technology.

In 2005-2006, several groups independently announced that the simulations 
of the inspiral black-hole binary merger are available
\cite{Pretorius,Goddard,UTB,LSU,PennState2006}. 
There are many implements for their successes, such as gauge conditions, coordinate selections, 
boundary treatments, singularity treatments, numerical discretization, 
and mesh refinements, together with the re-formulation of the Einstein equations 
which we will discuss here. 
More general and recent introductions to numerical
relativity are available, e.g. by 
Baumgarte-Shapiro (2003) \cite{reviewBS}, 
Alcubierre (2004) \cite{reviewAlcubierre}, 
Pretorius (2007) \cite{reviewPretorius}, and 
Bruegmann (2008) \cite{reviewBruegmann2008}.

The purpose of this article is to review the formulation problem in numerical relativity. 
This is one of the essential issues to achieve the long-term stable and accurate 
simulations of binary compact objects.  
Mathematically equivalent sets of evolution equations show different numerical 
stability in free evolution schemes.  
This had been the mystery for long time between the relativists, and many proposals and
trials were reported.  
After we review the problem from such a historical viewpoint, we will explain 
our systematic understanding using the constraint propagation equations; the evolution 
equations of the constraints which is supposed to be satisfied all through the time evolutions. 

The most numerical relativity groups today uses the so-called BSSN 
(Baumgarte-Shapiro-Shibata-Nakamura) equations, 
that is one of the modified form of the ADM (Arnowitt-Deser-Misner)
equations. 
We try to explain why these differences appear and also predict that more robust sets
of equations
exist together with actual numerical demonstrations.

\subsection{Formulation Problem in Numerical Relativity}

There are several different approaches to simulating the Einstein equations.  
Among them the most robust way, that we target in this article, is to 
apply 3+1  (space + time) decomposition of space-time.
This formulation was first given by  
Arnowitt, Deser and Misner (ADM) \cite{ADM} 
(we call the {\it original ADM system}, hereafter)
with the purpose of constructing a canonical formulation of the
Einstein equations to seek the quantum nature of space-time. 
In late 70s, when the numerical relativity started, this ADM formulation was
introduced by Smarr and York \cite{ADM-SmarrYork,ADM-York} in a slightly 
different notations which is taken
as the standard formulation between numerical relativists 
(so that we call the {\it standard ADM system}, hereafter). 

The ADM formulation divides the Einstein equations into the 
constraint equations and the evolution equations apparently, like the
Maxwell equations. 
Since the set of ADM equations form a first-class system, 
if we solved two constraint equations, 
the Hamiltonian (or energy) constraint and 
the momentum constraint equations for the 
initial data, then the evolution equations {\it theoretically guarantees}
the evolved data satisfy the constraint equations. 
This free-evolution approach is also the standard in numerical relativity. 
This is because solving the constraints (non-linear elliptic equations) is
numerically expensive, 
and because the free-evolution allows us to monitor the accuracy of
numerical evolution using the constraint equations. 

Up to the middle 90s, the ADM numerical relativity appealed great successes. 
For example, the formation of naked singularity from collisionless particles\cite{nakedsingularity} shows the unknown behavior of the strong gravity; 
the discovery of the critical behavior for a black-hole formation
\cite{criticalbehavior} open-doors the understanding of phase-transition 
nature in general relativity; the black-hole horizon dynamics
\cite{eventhorizon} realized the theoretical predictions.

\begin{figure}
\begin{center}
\includegraphics[width=50mm]{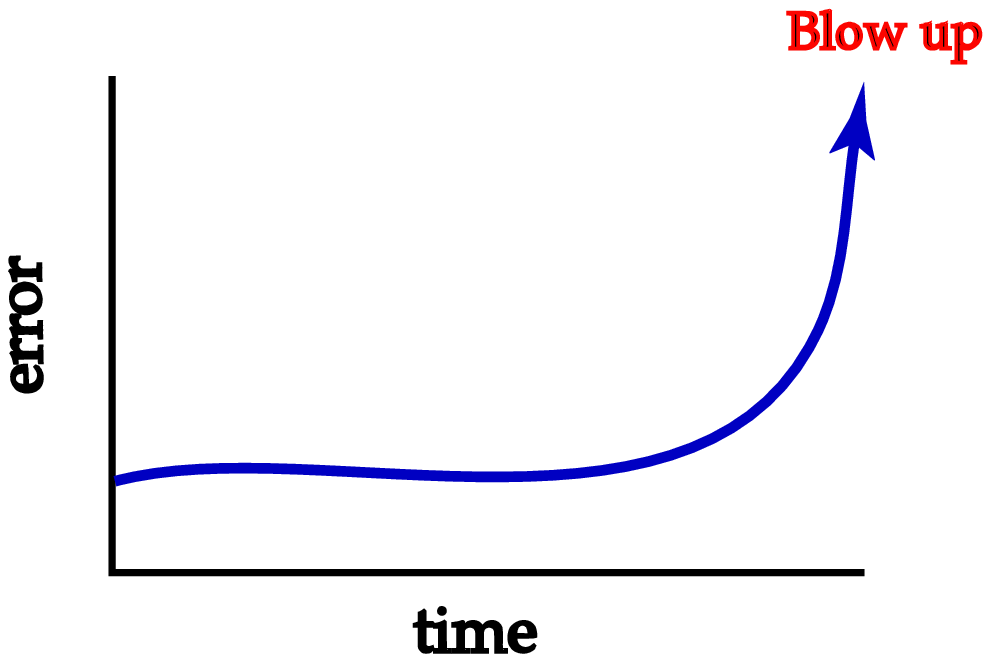}\\
\vspace{5mm}
\includegraphics[width=50mm]{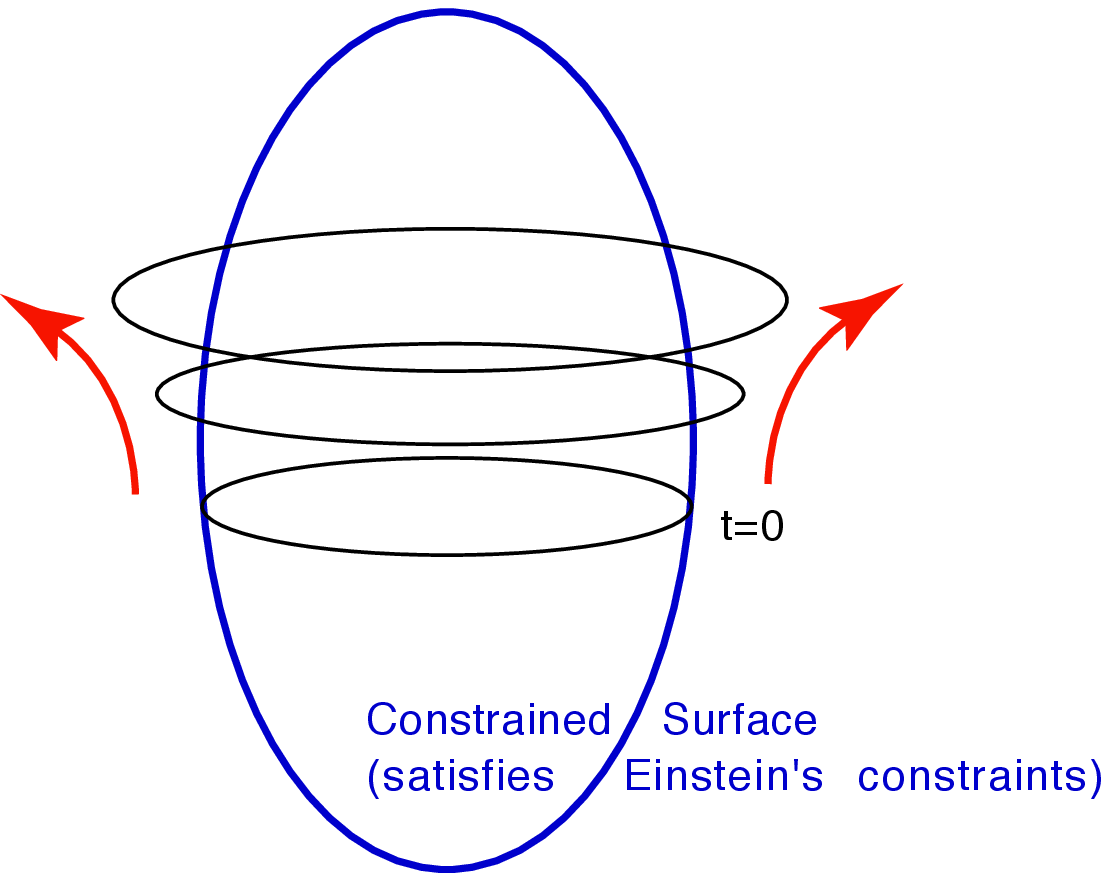}
\caption{Origin of the problem for numerical relativists: 
Numerical evolutions 
depart from the constraint surface. } \label{fig1}
\end{center}
\end{figure}

Nevertheless, when people try to make 
the long-term simulations such as coalescences
of neutron-star binaries and/or black-hole binaries for calculating
gravitational-wave form, 
numerical simulations were often interrupted by unexplained blow-ups
(Figure \ref{fig1}).  This was thought due to the lack of resolution, or  
inappropriate  gauge choice, or the particular numerical scheme which was applied.   
However, with the accumulation of experiences,  people have 
noticed the importance of the formulation of the evolution equations,
since there are apparent differences in numerical 
stability although the equations are mathematically equivalent. 

At this moment, 
there are three major ways to obtain longer time evolutions, which we 
describe in the next section. 
Of course, the ideas, procedures, and problems are mingled with each other. 
The purpose of this article is to review all three approaches and to 
introduce our idea to view them in a unified way. 
The author wrote a detail review of this topic in 2002 \cite{novabook}, 
and the present article includes an update in brief style.  


The word {\it stability} is used quite different ways in the community. 
\begin{itemize}
\item 
We mean by {\it numerical stability} a numerical simulation which continues
without any blow-ups and in which data remains on the constrained surface.  
\item {\it Mathematical stability} is defined in terms of the well-posedness
in the theory of partial differential equations, such that 
the norm of the variables is bounded by the initial
data. See eq. (\ref{energynorm}) and around.  
\item For numerical treatments, there is also another notion of 
{\it stability}, the stability
of finite differencing schemes.  This means that
numerical errors (truncation, round-off, etc) 
are not growing by evolution.  The evaluation is obtained using    
von Neumann's analysis. 
Lax's equivalence theorem says that if a numerical scheme is consistent
(converging to  the original equations in its continuum limit) and stable (no error
growing), then the simulation represents the right (converging) solution.  
See \cite{Choptuik91} for the Einstein equations.  
\end{itemize}

We follow the notations of that of MTW\cite{MTW}. 
We use
$\mu,\nu=0,\cdots,3$ and
$i,j=1,\cdots,3$ as space-time indices. The unit $c=1$ is applied.  
The discussion is mostly to the vacuum space-time, but the inclusion of matter is straightforward.

\section{The standard way and the three other roads}\label{sec2}
\subsection{Strategy 0: The ADM formulation}\label{secADM}
\subsubsection{The original ADM formulation}

\begin{figure}
\includegraphics[width=80mm]{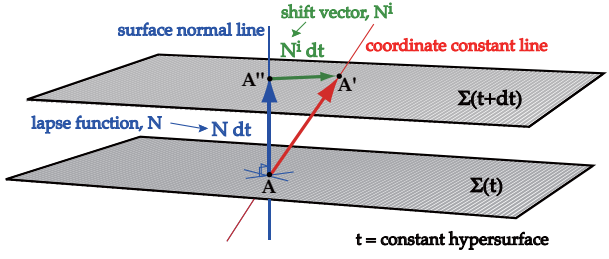}
\caption{Concept of time evolution of space-time: foliations of  
3-dimensional hypersurface. The lapse and shift functions are often denoted
$\alpha$ or $N$, and $\beta^i$ or $N^i$, respectively.}
\label{fig:ADMfoliation}
\end{figure}

The Arnowitt-Deser-Misner (ADM) formulation\cite{ADM} gave  
the fundamental idea of time evolution of space and time: such as 
foliations of 3-dimensional hypersurface (Figure \ref{fig:ADMfoliation}). 

The story begins by decomposing 4-dimensional space-time into 3 plus 1. 
The metric is expressed by 
\begin{eqnarray}
ds^2 &=& g_{\mu\nu} dx^\mu dx^\nu \nonumber \\
&=& - \alpha^2 dt^2+ 
\gamma_{ij}(dx^i + \beta^i dt)(dx^j + \beta^j dt), 
\end{eqnarray}
where $\alpha$ and $\beta_j$ are defined as 
$
\alpha \equiv 1 / \sqrt{-g^{00}}$ and $\beta_j \equiv g_{0j}, 
$
and called the lapse function and shift vector, respectively. 
The projection operator or the intrinsic 3-metric $g_{ij}$ is 
defined as 
$\gamma_{\mu\nu}=g_{\mu\nu}+n_\mu n_\nu$, where
$n_\mu=(-\alpha, 0,0,0)$ [and $n^\mu=g^{\mu\nu}n_\nu=
(1/\alpha, -\beta^i/\alpha)$] is the unit normal vector of the spacelike 
hypersurface, $\Sigma$ (see Figure \ref{fig:ADMfoliation}). 
By introducing the extrinsic curvature,  
\begin{eqnarray}
K_{ij}&=& - \frac{1}{2} \pounds_n \gamma_{ij}, 
\end{eqnarray}
and using the Gauss-Codacci relation, the Hamiltonian density of the Einstein equations can be written as 
\begin{equation}
{\cal H}_{GR} = \pi^{ij}\dot{\gamma}_{ij} - {\cal L}, 
\end{equation}
where
\begin{equation}
{\cal L}=\sqrt{-g}R=\alpha\sqrt{\gamma}[{}^{(3)\!}R - K^2 + K_{ij}K^{ij}], 
\end{equation}
where $\pi^{ij}$ is the canonically conjugate momentum to $\gamma_{ij}$,  
\begin{equation}
\pi^{ij}={\partial {\cal L} \over \partial \dot{\gamma}_{ij}}
= - \sqrt{\gamma} (K^{ij} - K \gamma^{ij}), 
\end{equation}
omitting the boundary terms. 
The variation of ${\cal H}_{GR}$ with
respect to $\alpha$ and $\beta_i$ yields the constraints, and 
the dynamical equations are given by 
$\dot{\gamma}_{ij} = {\delta {\cal H}_{GR} \over \delta \pi^{ij} } $ and 
$\dot{\pi}^{ij} = -{\delta {\cal H}_{GR} \over \delta h_{ij} }$.

\subsubsection{The standard ADM formulation}
In the version of Smarr and York\cite{ADM-SmarrYork,ADM-York}, $K_{ij}$ was used as a fundamental variable
instead of the conjugate momentum $\pi^{ij}$. 
The set of equation is summarized as follows: \\

\noindent
$\Box ${\it The Standard ADM formulation \cite{ADM-SmarrYork,ADM-York}} \\
The fundamental dynamical variables are $(\gamma_{ij}, K_{ij})$, 
the three-metric and extrinsic curvature.  
The three-hypersurface $\Sigma$ is foliated with gauge functions, 
($\alpha, \beta^i$), the lapse and shift vector. 
\begin{itemize}
\item 
The evolution equations:
\begin{eqnarray}
{\partial_t} \gamma_{ij} &=& 
-2\alpha K_{ij}+D_i\beta_j+D_j\beta_i, 
 \label{adm_evo1}
 \\
{\partial_t} K_{ij} &=& 
\alpha ~^{(3)\!} R_{ij}+\alpha K K_{ij}-2\alpha K_{ik}{K^k}_j 
-D_iD_j \alpha 
\nonumber \\&& 
+(D_i \beta^k) K_{kj} +(D_j \beta^k) K_{ki} 
+\beta^k D_k K_{ij} 
\label{adm_evo2}
\end{eqnarray}
where $K=K^i{}_i$, and $~^{(3)\!}R_{ij}$ and $D_i$ denote
 three-dimensional Ricci curvature, 
and a covariant derivative on the three-surface, respectively.  
\item
Constraint equations:
\begin{eqnarray}
{\cal H}^{ADM} &:=&
~^{(3)\!}R+ K^2 -K_{ij}K^{ij} \approx 0,
\label{admCH} \\
{\cal M}^{ADM}_i &:=&
D_j K^j{}_i-D_i K  
\approx 0,
\label{admCM}
\end{eqnarray}
where $~^{(3)\!}R=^{(3)\!}R^i{}_i$: these
are called the Hamiltonian (or energy) and momentum
constraint equations, respectively. $\qquad \Box$
\end{itemize}

The formulation has 12 first-order dynamical variables 
($\gamma_{ij}, K_{ij}$), with 4 freedom of gauge choice ($\alpha, \beta_i$)
 and with 4 constraint equations,  (\ref{admCH}) and (\ref{admCM}). 
The rest freedom expresses 2 modes of gravitational waves. 

We remark that there is one replacement in (\ref{adm_evo2}) using (\ref{admCH})
in the process of conversion from
the original ADM to the standard ADM equations. 
This is the key issue in the later discussion, and we shall be back this point 
in \S \ref{secADJADM}.

\begin{figure*}
\begin{center}
\includegraphics[width=120mm]{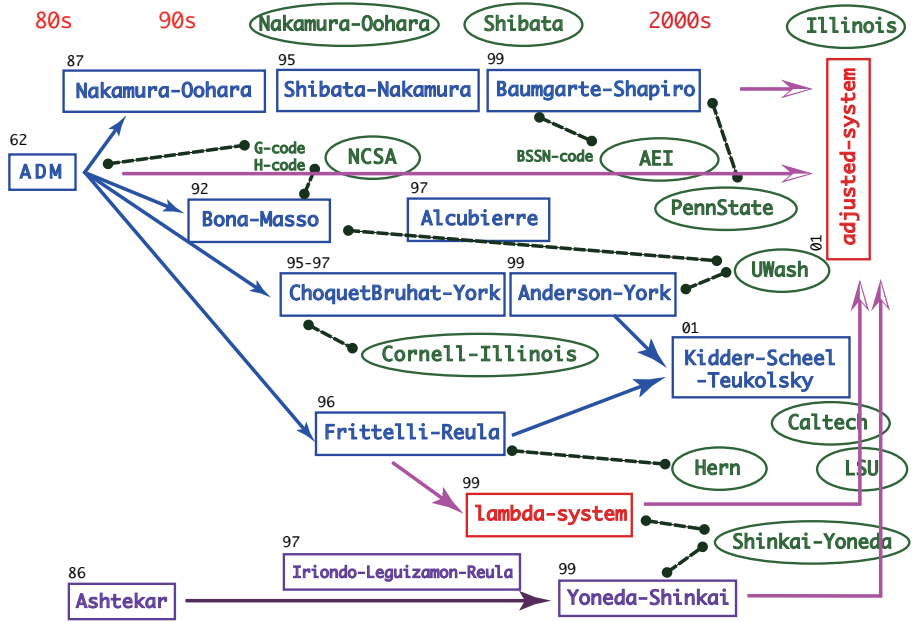}
\caption{Chronological table of formulations and their numerical tests ($\sim$ 2001). 
Boxed ones are of proposals of formulation, circled ones are related
numerical experiments.  Please refer Table 1 in \cite{novabook} for
each references.}
\label{fig3history1}
\end{center}
\end{figure*}

The constraint propagation equations,
which are the time evolution equations
of the Hamiltonian constraint (\ref{admCH}) and
the momentum constraints (\ref{admCM}), can be written as follows: 

\noindent
$\Box ${\it Constraint Propagations of the Standard ADM:} \\
\begin{eqnarray}
\partial_t {\cal H}&=&
\beta^j (\partial_j {\cal H})
+2\alpha K{\cal H}
-2\alpha \gamma^{ij}(\partial_i {\cal M}_j)
\nonumber \\ &&
+\alpha(\partial_l \gamma_{mk})
   (2\gamma^{ml}\gamma^{kj}-\gamma^{mk}\gamma^{lj}){\cal M}_j
\nonumber \\ &&
-4\gamma^{ij} (\partial_j\alpha){\cal M}_i,
\label{CHproADM}
\\
\partial_t {\cal M}_i&=&
-(1/2)\alpha (\partial_i {\cal H})
-(\partial_i\alpha){\cal H}
+\beta^j (\partial_j {\cal M}_i)
+\alpha K {\cal M}_i
\nonumber \\ &&
-\beta^k\gamma^{jl}(\partial_i\gamma_{lk}){\cal M}_j
+(\partial_i\beta_k)\gamma^{kj}{\cal M}_j.
\qquad \Box
\label{CMproADM}
\end{eqnarray}
From these equations, we know that  
{\it if the constraints are satisfied 
on the initial slice $\Sigma$, then 
the constraints are satisfied throughout evolution}.
The normal numerical scheme is to solve the elliptic constraints
for preparing the initial data, and to apply the free evolution (solving only 
the evolution equations).  The constraints are used to monitor the accuracy of 
simulations. 

The ADM formulation was the standard formulation for numerical relativity 
up to the middle 90s.  
Numerous successful simulations were obtained for the problems of
gravitational collapse, critical behavior, cosmology, and so on.  
However,  stability problems have arisen for the
simulations such as the
gravitational radiation from compact binary coalescence, because the
models require quite a long-term time evolution. 

\begin{figure*}
\begin{center}
\includegraphics[width=120mm]{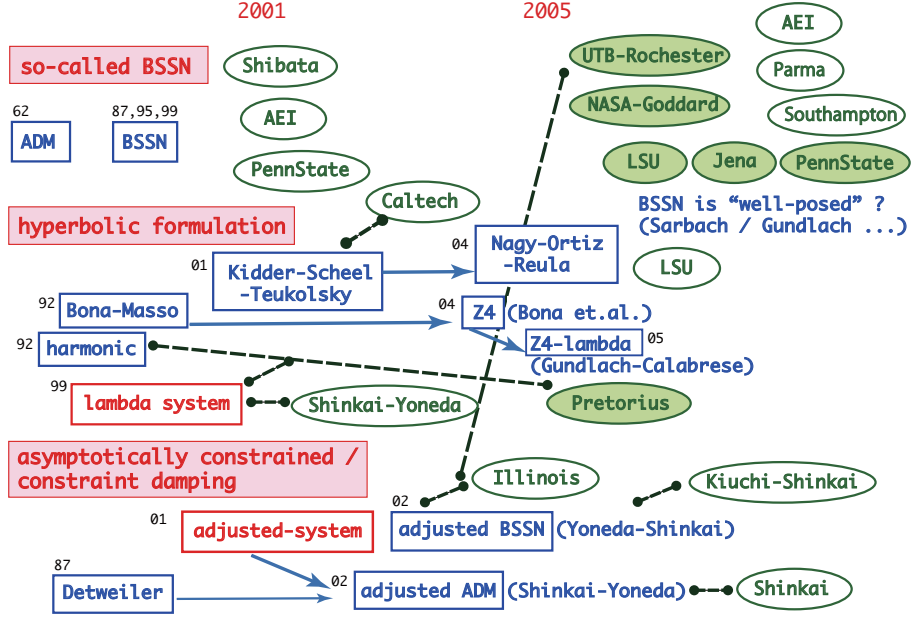}
\caption{Chronological table of formulations and their numerical tests
(2001 $\sim$).  
Boxed ones are of proposals of formulation, circled ones are related
numerical experiments.  }
\label{fig4history2}
\end{center}
\end{figure*}

The origin of the problem was that the above  statement in {\it Italics}
is true in principle, but is not always true in numerical applications. 
A long history of trial  and error  began in the early 90s. 
From the next subsection we shall 
look back of them by summarizing  ``three strategies".
We then unify these three roads as ``adjusted systems", and as its by-product 
we show 
that the standard ADM equations {\it has} a  
constraint violating mode in its constraint propagation equations
even for a single black-hole (Schwarzschild) spacetime \cite{adjADMsch}. 
Figure \ref{fig3history1} and \ref{fig4history2} are chronological maps of the researches.

\subsection{Strategy 1: Modified ADM formulation by Nakamura et al 
(The BSSN formulation)
} \label{secBSSN}
Up to now, the most widely used formulation 
for large scale numerical simulations is 
a modified ADM system, which is now often cited as the 
Baumgarte-Shapiro-Shibata-Nakamura (BSSN) formulation. 
This re-formulation was first introduced by Nakamura {\it et al.} 
\cite{SN87,SN89,SN}. 
The usefulness of this re-formulation was re-introduced by 
Baumgarte and Shapiro \cite{BS}, then was 
confirmed by other groups to show a long-term
stable numerical evolution 
\cite{potsdam9908,potsdam0003}.

\subsubsection{Basic variables and equations}
The widely used notation\cite{BS} introduces the variables 
($\varphi,\tilde{\gamma}_{ij}$,$K$,$\tilde{A}_{ij}$,$\tilde{\Gamma}^i$) 
instead of ($\gamma_{ij}$,$K_{ij}$), where
\begin{eqnarray}
\varphi &=& {1\over 12}\log ({\rm det}\gamma_{ij}), \\
\label{BSSNval1} 
\tilde{\gamma}_{ij} &=&
e^{-4\varphi}\gamma_{ij}, 
\label{BSSNval2} \\
K  &=& \gamma^{ij}K_{ij}, \label{BSSNval3} \\
\tilde{A}_{ij} &=&
e^{-4\varphi}(K_{ij} - (1/3)\gamma_{ij}K), \label{BSSNval4} 
\\ 
\tilde{\Gamma}^i &=&
\tilde{\Gamma}^i_{jk}\tilde{\gamma}^{jk}.
\label{BSSNval5} 
\end{eqnarray}
The new variable $\tilde{\Gamma}^i$ is introduced in order to 
calculate Ricci curvature more accurately.  
In BSSN formulation, Ricci curvature is not calculated as 
$R^{ADM}_{ij} 
= 
\partial_k \Gamma^k_{ij} 
-\partial_i\Gamma^k_{kj} 
+\Gamma^l_{ij}\Gamma^k_{lk} 
-\Gamma^l_{kj}\Gamma^k_{li}, 
$
but as $R^{BSSN}_{ij} =
R^\varphi_{ij}+\tilde R_{ij}$, where the first term includes
the conformal factor $\varphi$ while the second term does not. 
These are approximately equivalent, but $R^{BSSN}_{ij}$ does have 
wave operator apparently in the flat background limit, so that 
we can expect more natural wave propagation behavior. 

Additionally, the BSSN requires us to impose the conformal factor as 
$
\tilde{\gamma}(:={\rm det} \tilde{\gamma}_{ij})=1, 
$
during evolution.  This is a kind of definition, but can also be 
treated as a constraint. 


\noindent
$\Box ${\it The BSSN formulation \cite{SN87,SN89,SN,BS}:}\\
The fundamental dynamical variables are
($\varphi,\tilde{\gamma}_{ij}$,$K$,$\tilde{A}_{ij}$,$\tilde{\Gamma}^i$).
\\
The three-hypersurface $\Sigma$ is foliated with gauge functions, 
($\alpha, \beta^i$), the lapse and shift vector. 
\begin{itemize}
\item 
The evolution equations:
\begin{eqnarray}
\partial_t^B \varphi 
&=& 
-(1/6)\alpha K+(1/6)\beta^i(\partial_i\varphi)+(\partial_i\beta^i), 
\label{BSSNeqmPHI} 
\\ 
\partial_t^B \tilde{\gamma}_{ij} 
&=& 
-2\alpha\tilde{A}_{ij} 
+\tilde{\gamma}_{ik}(\partial_j\beta^k) 
+\tilde{\gamma}_{jk}(\partial_i\beta^k) 
\nonumber \\ && 
-(2/3)\tilde{\gamma}_{ij}(\partial_k\beta^k) 
+\beta^k(\partial_k\tilde{\gamma}_{ij}), 
\label{BSSNeqmtgamma} 
\\ 
\partial_t^B K 
&=& 
-D^iD_i\alpha 
+\alpha \tilde{A}_{ij}\tilde{A}^{ij} 
+(1/3) \alpha K^2 
+\beta^i (\partial_i K), 
\label{BSSNeqmK} 
\\ 
\partial_t^B \tilde{A}_{ij} 
&=& 
-e^{-4\varphi}(D_iD_j\alpha)^{TF} 
+e^{-4\varphi} \alpha (R^{BSSN}_{ij})^{TF} 
\nonumber \\&& 
+\alpha K\tilde{A}_{ij} 
-2\alpha \tilde{A}_{ik}\tilde{A}^k{}_j 
+(\partial_i\beta^k)\tilde{A}_{kj} 
+(\partial_j\beta^k)\tilde{A}_{ki} 
\nonumber \\&& 
-(2/3)(\partial_k\beta^k)\tilde{A}_{ij} 
+\beta^k(\partial_k \tilde{A}_{ij}), 
\label{BSSNeqmTA} 
\\ 
\partial_t^B \tilde{\Gamma}^i &=& 
-2(\partial_j\alpha)\tilde{A}^{ij} 
+2\alpha 
\big(\tilde{\Gamma}^i_{jk}\tilde{A}^{kj} 
-(2/3)\tilde{\gamma}^{ij}(\partial_j K) 
\nonumber \\&& 
+6\tilde{A}^{ij}(\partial_j\varphi) 
\big) 
-\partial_j 
\big( \beta^k(\partial_k\tilde{\gamma}^{ij}) 
-\tilde{\gamma}^{kj}(\partial_k\beta^{i}) 
\nonumber \\&& 
-\tilde{\gamma}^{ki}(\partial_k\beta^{j}) 
+(2/3)\tilde{\gamma}^{ij}(\partial_k\beta^k) 
\big). 
\label{BSSNeqmTG} 
\end{eqnarray}
\item
Constraint equations:
\begin{eqnarray}
{\cal H}^{BSSN} 
&=& 
R^{BSSN}+K^2-K_{ij}K^{ij}, 
\label{BSSNconstraintH} 
\\ 
{\cal M}^{BSSN}_i 
&=& 
{\cal M}^{ADM}_i,  \label{BSSNconstraintM} 
\\
{\cal G}^i &=& \tilde{\Gamma}^i-\tilde{\gamma}^{jk} 
\tilde{\Gamma}^i_{jk}, \label{BSSNconstraintG} 
\\ 
{\cal A}&=&\tilde{A}_{ij}\tilde{\gamma}^{ij}, \label{BSSNconstraintA} 
\\ 
{\cal S} &=& 
\tilde{\gamma}-1. \qquad \Box
\label{BSSNconstraintS} 
\end{eqnarray}
\end{itemize}
(\ref{BSSNconstraintH}) and (\ref{BSSNconstraintM}) are 
the Hamiltonian and momentum constraints 
(the ``kinematic" constraints), while the latter three are 
 ``algebraic" constraints due to the requirements of BSSN variables.

\subsubsection{Remarks, Pros and Cons}\label{sec:BSSNproscons}
Why BSSN is better than the standard ADM? 
Together with numerical comparisons with the 
standard ADM case\cite{potsdam0003}, 
this question has been studied by many groups using different approaches. 
\begin{itemize}
\item 
Using numerical test evolutions, 
Alcubierre  {\it et al.}  \cite{potsdam9908} found that 
the essential improvement is in 
the process of replacing terms by the momentum constraints. 
They also pointed out that 
the eigenvalues of BSSN {\it evolution equations} have fewer 
``zero eigenvalues" 
than those of ADM, and they conjectured that the instability 
might be caused by these 
``zero eigenvalues". 
\item 
Miller\cite{Miller} reported that  
BSSN has a wider range of 
parameters that gives us stable evolutions 
in von Neumann's stability analysis. 
\item 
An effort was made to understand the advantage of BSSN from the point
of hyperbolization of the equations in its linearized
limit \cite{potsdam9908,LSU-BSSN}, or with a particular combination of 
slicing conditions plus auxiliary variables\cite{HeyerSarbach}.
If we define the 2nd order symmetric hyperbolic form, then the 
principal part of BSSN can be one of them\cite{Gundlach0406}. 
\end{itemize}

As we discussed in \cite{adjBSSN}, the stability of the
BSSN formulation is due not only to the introductions of 
new variables, but also to the replacement of terms in the evolution 
equations using the constraints.  Further, we can show several 
additional adjustments to the BSSN equations which 
give us more stable numerical simulations.  
We will devote \S \ref{sec3} for the fundamental idea.

The current binary black-hole simulations apply the BSSN formulations
with several implementations. For example,  
\begin{itemize}
\item[tip-1]
Alcubierre  {\it et al.}  \cite{potsdam0003} reported that 
trace-out $A_{ij}$ technique at every time-step helps the stability. 
\item[tip-2]
Campanelli  {\it et al.}  \cite{UTB} reported that in their code 
$\tilde{\Gamma}^i$ is replaced by $-\partial_j\tilde{\gamma}^{ij}$
where it is not differentiated. 
\item[tip-3]
Baker  {\it et al.}  \cite{Goddard} modified $\tilde{\Gamma}^i$-equation (\ref{BSSNeqmTG})
as suggested by
Yo  {\it et al.}   \cite{YBS}. 
\end{itemize}
These technical tips are again explained using the constraint propagation 
analysis as we will come back in \S \ref{secADJBSSNa}. 

These studies provide sort of evidences regarding
the advantage of BSSN, while it is also shown 
an example of an ill-posed solution in BSSN (as well in ADM) 
by Frittelli and Gomez \cite{FrittelliGomez}. 
Recently, the popular combination, BSSN with Bona-Masso type slicing condition, 
is investigated in particular. Among then, Garfinkle  {\it et al.}  
\cite{GGH0707} 
speculated that the reason of
gauge shocks are missing in the current 3-dimensional black-hole simulations 
is simply because the lack of resolution.


\subsection{Strategy 2: Hyperbolic re-formulations} 
\label{secHYP}
\subsubsection{Definitions, properties, mathematical backgrounds}

The second effort to re-formulate the Einstein equations is to make 
the evolution equations reveal a
first-order hyperbolic form explicitly. 
This is motivated by the expectation that the symmetric hyperbolic 
system has well-posed
properties in its Cauchy treatment in many systems and also that 
the boundary treatment can be
improved if we know the characteristic speed of the system.  

\noindent
$\Box ${\it Hyperbolic formulations} \\ 
We say that the system is {\it a first-order (quasi-linear)
partial differential equation system}, 
if a certain set of
(complex-valued) variables $u_\alpha$ $(\alpha=1,\cdots, n)$
forms
\begin{equation}
\partial_t u_\alpha
= {\cal M}^{l}{}^{\beta}{}_\alpha (u) \, \partial_{l} u_\beta
+{\cal N}_\alpha(u),
\label{def}
\end{equation}
where ${\cal M}$ (the characteristic matrix) and
${\cal N}$ are functions of $u$
but do not include any derivatives of $u$.  Further we say the system
is 
\begin{itemize}
\item {\it a weakly hyperbolic system}, 
if all the eigenvalues of the characteristic matrix are real. 
\item {\it a strongly hyperbolic system}  
(or a diagonalizable / symmetrizable hyperbolic  system),
if the characteristic matrix is
diagonalizable (has a complete set of eigenvectors) and has all real eigenvalues. 
\item {\it a symmetric hyperbolic system}, 
if the characteristic matrix is a Hermitian matrix.  $\qquad\Box$
\end{itemize}

Writing the system in a hyperbolic form is a
quite useful 
step in proving that the system is {\it well-posed}.  
The mathematical well-posedness of the system means
($1^\circ$) local existence (of at least one solution $u$), 
($2^\circ$) uniqueness (i.e., at most solutions),  and 
($3^\circ$) stability (or continuous dependence of
solutions $\{ u \}$ on the Cauchy data) 
of the solutions. 
The resultant statement expresses the existence of the energy inequality 
on its norm, 
\begin{eqnarray}
&& || u(t) || \leq e^{\alpha \tau} || u(t=0) ||,  \nonumber \\
&&\quad \mbox{\rm where~} 0 < \tau < t, \quad
\alpha={\rm const.} 
\label{energynorm}
\end{eqnarray}
This indicates that the norm of $u(t)$ is bounded by a certain function and
the initial norm.  Remark that this mathematical bounds 
does not mean that the norm $u(t)$ decreases along the time evolution. 

The inclusion relation of the hyperbolicities  is,  
\begin{eqnarray}
\mbox{\rm symmetric~hyperbolic} &\subset&
\mbox{\rm strongly~hyperbolic} \nonumber \\ &\subset&
\mbox{\rm weakly~hyperbolic}.
\end{eqnarray}
The Cauchy problem under weak hyperbolicity is not,
  in general, $C^\infty$ well-posed.
At the strongly hyperbolic level,
we can prove the finiteness of the energy norm
if the characteristic matrix is independent of $u$
(cf \cite{Stewart}), that is one step definitely advanced over 
a weakly hyperbolic form.
Similarly, 
the well-posedness of the symmetric hyperbolic is guaranteed 
if the characteristic matrix is independent of $u$,
while if it depends on $u$ we have only limited proofs for the
well-posedness.

{}From the point of numerical applications, 
to hyperbolize the evolution equations 
is quite attractive, 
not only for its mathematically well-posed features. 
The expected additional advantages are the following.
\begin{itemize}
\item[(a)] 
It is well known that a certain flux conservative hyperbolic
system is taken as an essential formulation in the
computational Newtonian hydrodynamics when we control shock
wave formations due to matter. 
\item[(b)] 
The characteristic speed (eigenvalues of the principal matrix) 
is supposed to be the
propagation speed of the information in that system. 
Therefore it is naturally imagined that these magnitudes are
equivalent to the physical information speed of the model to 
be simulated. 
\item[(c)] 
The existence of the characteristic speed of the system 
 is expected to give us 
an improved treatment of the numerical boundary, and/or to 
give us a new well-defined 
 Cauchy problem within a finite region 
(the so-called initial boundary value problem; IBVP). 
\end{itemize}
These statements sound reasonable, but have not yet been generally 
confirmed in actual numerical simulations in general relativity.

\subsubsection{Hyperbolic formulations of the Einstein equations}
Most physical systems
can be expressed as symmetric hyperbolic systems.
In order to prove that the Einstein's theory is a well-posed system, 
to hyperbolize the Einstein equations is a long-standing research area 
in mathematical relativity.
  
The standard ADM system does not form a first
order hyperbolic system.
This can be seen immediately from the fact that the
ADM evolution equation (\ref{adm_evo2}) has Ricci curvature in RHS.
This is also the common fact to the BSSN formulation. 

So far, several first order hyperbolic systems of the Einstein equation
have been proposed.  
In constructing
hyperbolic systems, the essential procedures are 
(1$^\circ$)
to introduce new variables, normally the spatially derivatived metric, 
(2$^\circ$)
to adjust equations using constraints.  
Occasionally, (3$^\circ$) to restrict the gauge conditions, and/or (4$^\circ$)
to rescale some variables. 
Due to process (1$^\circ$), 
the number of fundamental
dynamical variables is always larger than that of ADM. 

Due to the limitation of space, we can only list 
several hyperbolic systems of the Einstein
equations:  
\begin{itemize}
\item The Bona-Mass\'o formulation \cite{BM,BMSS} 
\item The Einstein-Ricci system \cite{CY9506071,CY9506072} / 
Einstein-Bianchi system \cite{CY9710041} 
\item The Einstein-Christoffel system \cite{AY} 
\item The Ashtekar formulation \cite{Ashtekar,ysPRL}  
\item The Frittelli-Reula formulation \cite{FR96,Stewart}  
\item The Conformal Field equations \cite{FriedrichCFE} 
\item The Bardeen-Buchman system \cite{BB}
\item The Kidder-Scheel-Teukolsky (KST) formulation \cite{KST} 
\item The Alekseenko-Arnold system \cite{AA2003} 
\item The general-covariant Z4 system \cite{Z4} 
\item The Nagy-Ortiz-Reula (NOR) formulation \cite{NOR} 
\item The Weyl system \cite{FriedrichWeyl,FrauendienerVogel} 
\end{itemize}
Note that there is no apparent differences between the word `formulation' and 
`system' here. 
\subsubsection{Numerical tests} \label{sec:hypnumeric}
When we discuss hyperbolic systems in the context of numerical 
stability, the following questions should be considered:  

\begin{itemize}
\item[Q]
From the point of the set of evolution equations, 
does hyperbolization actually contribute to numerical accuracy
and stability?   
Under what conditions/situations will the advantages of 
hyperbolic formulation be observed?
\end{itemize}
Unfortunately, we do not have conclusive answers to these questions, but many 
experiences are being accumulated. 
Several earlier numerical comparisons reported the stability of  
hyperbolic formulations \cite{BMSS,cactus1,SBCSThyper,SBCST98}.  
But we have to remember that 
this statement went against the standard ADM formulation. 

These partial numerical successes encouraged the community to 
formulate various hyperbolic systems.  
However, several numerical experiments also indicate that this direction
is not a complete success. 

\begin{itemize}
\item Above earlier numerical successes were also terminated with blow-ups.  
\item If the gauge functions are evolved according to the hyperbolic 
equations, then their finite propagation speeds may cause pathological 
shock formations in simulations \cite{Alcubierre,AM}.  
\item 
There are no
drastic differences in the evolution properties {\it between} 
hyperbolic systems (weakly, strongly and symmetric hyperbolicity)
by systematic numerical studies 
by Hern \cite{HernPHD} based on Frittelli-Reula formulation \cite{FR96}, and
by the authors \cite{ronbun1} based on Ashtekar's formulation 
\cite{Ashtekar,ysPRL}.
\item Proposed symmetric hyperbolic systems were not always the best ones 
for numerical evolution.  People are normally still required to re-formulate
them for suitable evolution.  Such efforts are seen in the applications of
the Einstein-Ricci system \cite{SBCST98},  
the Einstein-Christoffel system \cite{BB},  
and so on.
\item If we can erase the non-principal part by suitable re-definition of variables (as is 
in the KST formulation)\cite{KST},  then we can see the growth of energy norm (\ref{energynorm})
in numerical evolution as theoretically predicted \cite{LindblomScheel,LSU-KST}. 
We then see a certain differences in the long-term
convergence features between weakly and strongly hyperbolic systems. 
\end{itemize}
Of course, these statements only casted on a particular formulation, and therefore
we have to be careful not to over-emphasize the results.

\subsubsection{Remarks} \label{sec:hypRemark}

 In order to 
figure out the reasons for the above objections, 
it is worth stating the following cautions:

\begin{itemize}
\item[(a)]
Rigorous mathematical proofs of well-posedness of PDE are
mostly for simple symmetric or strongly hyperbolic systems.  
If the matrix components or coefficients depend on dynamical variables
(as in all any versions of hyperbolized Einstein equations), 
almost nothing was proved in more general situations. 
\item[(b)]
The statement of ``stability" in the discussion of well-posedness 
refers to the bounded growth of the norm, (\ref{energynorm}), and does not indicate
a decay of the norm in time evolution. 
\item[(c)]
The discussion of hyperbolicity only uses
the characteristic part of the evolution equations, and ignores the rest. 
\end{itemize}

We think the origin of confusion  in the community results from over-expectation
on the above issues.  Mostly, point (c) is the biggest problem. 
The above numerical claims from Ashtekar\cite{ronbun1,ronbun2} and 
Frittelli-Reula \cite{HernPHD} formulations
were mostly due to the contribution (or interposition) 
of non-principal parts in evolution. 
Regarding this issue, 
the KST formulation finally opens the door.  
KST's
``kinematic" parameters enable us to reduce the non-principal part, so that
numerical experiments are hopefully expected to represent predicted
evolution features from PDE theories.  
At this moment, the agreement between numerical behavior and 
theoretical prediction is not yet perfect but close 
\cite{LindblomScheel}. 

If further studies reveal the direct correspondences between theories and numerical
results, then the direction of hyperbolization will remain as the 
essential approach in numerical relativity, and the related IBVP researches 
\cite{Stewart,IBVP-FN,IBVP-BS,IBVP-KRSW,IBVP-RRS}
will become a main research subject in the future. 
Meanwhile, 
it will be useful if we have an alternative procedure
to predict stability including the effects of the 
non-principal parts of the equations. 
Our proposal of adjusted system in the next subsection may be one of them. 

\subsection{Strategy 3: Asymptotically constrained systems}  
\label{secASYMPT}

\begin{figure}[t]
\includegraphics[width=50mm]{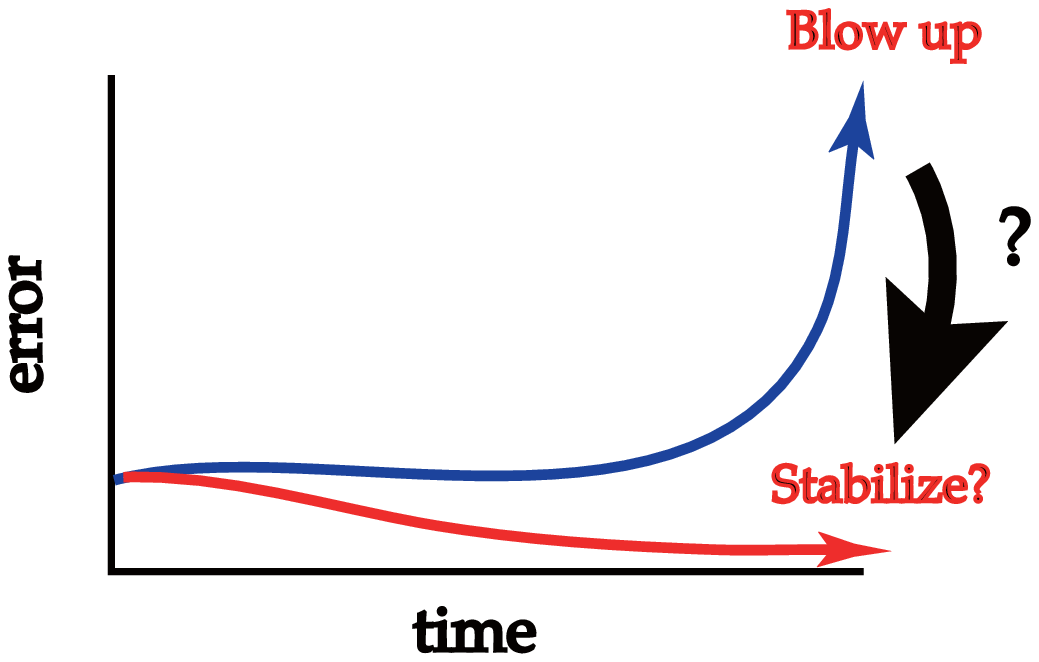}\\
\vspace{8mm}
\includegraphics[width=50mm]{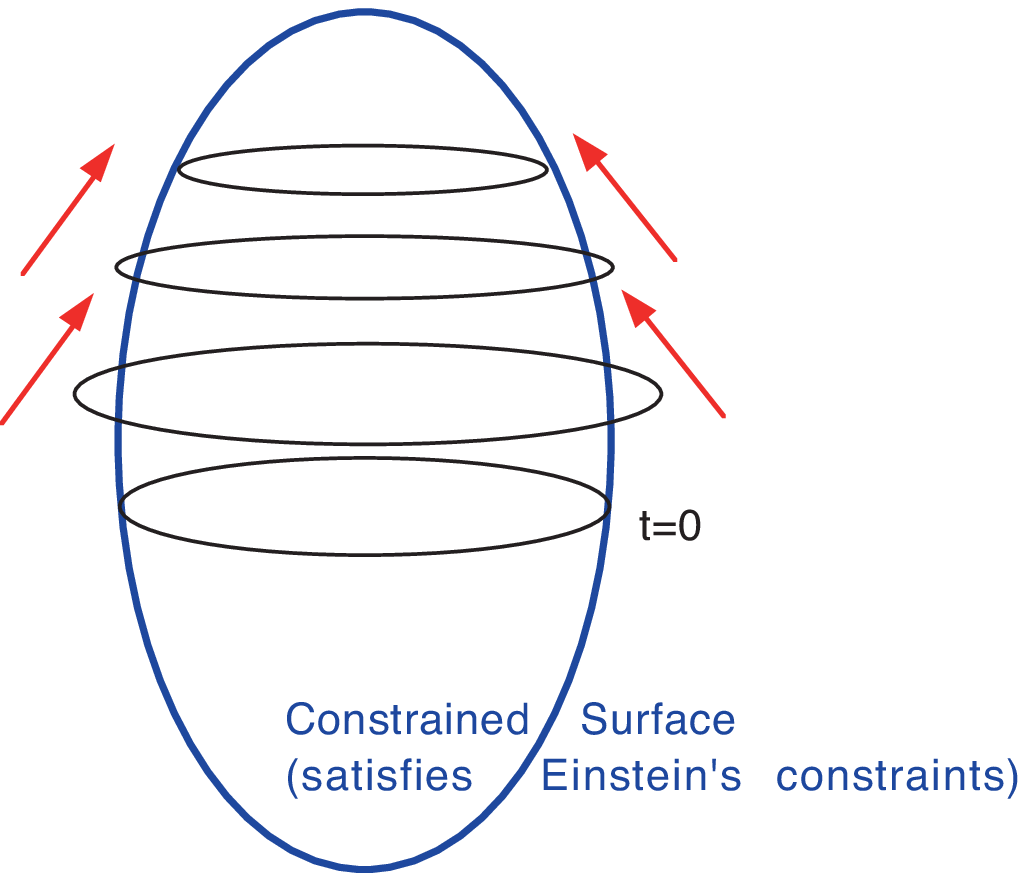}
\caption{The idea of the ``Asymptotically Constrained System". } 
\label{fig:attractor}
\end{figure}

The third strategy is to construct a robust system against 
the violation of constraints,
such that the constraint surface is an attractor (Figure \ref{fig:attractor}).  
The idea was first proposed as ``$\lambda$-system" 
by Brodbeck  {\it et al.} \cite{BFHR}, and then developed in more general situations 
as ``adjusted system" by the authors \cite{ronbun2}.  

\subsubsection{The ``$\lambda$-system"} 
\label{sec:lambdasystem}
Brodbeck  {\it et al.}  \cite{BFHR} proposed a system which has additional variables
$\lambda$ that obey artificial dissipative equations.  
The variable $\lambda$s are supposed to indicate the violation of 
constraints and the target of the system is to get $\lambda=0$ as its
attractor.  

\noindent
$\Box ${\it The ``$\lambda$-system" (Brodbeck et al.) 
\cite{BFHR}:} \\ 
For a symmetric hyperbolic system, 
add additional variables $\lambda$ and 
artificial force to reduce the violation of
constraints. 
The procedure is as follows: 
\begin{enumerate}
\item {Prepare a symmetric hyperbolic evolution system}
\begin{equation}
\partial_t u=M \partial_i u+N
\end{equation}
\item 
{Introduce $\lambda$ as an indicator of violation of }
{constraint which obeys dissipative eqs. of motion }
\begin{equation}
\partial_t \lambda=\alpha C-\beta \lambda, (\alpha \neq 0, \beta>0)
\end{equation}
\item {Take a set of $(u,\lambda)$ as dynamical variables }
\begin{equation}
\partial_t
\left(\begin{matrix} u \\ \lambda \end{matrix}\right)
\simeq
\left(\begin{matrix}A & 0 \\ F & 0 \end{matrix}\right)
\partial_i 
\left(\begin{matrix} u \\ \lambda \end{matrix} \right)
\end{equation}
\item 
{Modify evolution eqs so as to form  } 
{a
symmetric hyperbolic system
} 
\begin{equation}
\partial_t
\left(\begin{matrix}  u \\ \lambda \end{matrix} \right)
=
\left(\begin{matrix} A & \bar{F} \\ F& 0\end{matrix}\right)
\partial_i 
\left(\begin{matrix} u \\ \lambda \end{matrix} \right)
\qquad \Box
\end{equation}
\end{enumerate}
Since the total system is designed 
to have symmetric hyperbolicity,  
the evolution is supposed to be unique.  
Brodbeck {\it et al.} showed analytically that such a decay of $\lambda$s can be seen
for sufficiently small $\lambda (>0)$ 
with a choice of appropriate combinations of $\alpha$s and $\beta$s.

Brodbeck {\it et al.}  presented a set of equations
based on Frittelli-Reula's symmetric hyperbolic formulation \cite{FR96}. 
The version of Ashtekar's variables was presented by the authors 
\cite{SY-asympAsh}
for controlling the constraints or reality conditions or both. 
The numerical tests of both the Maxwell-$\lambda$-system and 
the Ashtekar-$\lambda$-system were performed \cite{ronbun2}, and 
 confirmed to work as expected. 
 The $\lambda$-system version of the general-covariant Z4 system \cite{Z4} 
 is also presented \cite{Z4lambda}. 
Pretorius \cite{Pretorius} applied this ``constraint-damping" idea in
his harmonic system to perform his binary black-hole merger simulations. 
 
Although it is questionable whether the
recovered solution is true evolution or 
not \cite{SiebelHuebner}, we think 
the idea is quite attractive. 
To enforce the decay of errors in its initial 
perturbative stage seems the key 
to the next improvements, which are also
 developed in the next section on ``adjusted systems". 

However, 
there is a high price to pay for constructing a $\lambda$-system.
The $\lambda$-system can not be introduced generally, because
(i) the construction of $\lambda$-system requires the original evolution 
equations to have a symmetric hyperbolic form, which is quite restrictive for
the Einstein equations, 
(ii) the final system requires many additional variables
and we also need to evaluate all the constraint equations at every time step,
which is a hard task in computation.  
Moreover, 
(iii) it is not clear that the $\lambda$-system is robust enough for 
non-linear violation of constraints, or that $\lambda$-system 
can control constraints
which do not have any spatial differential terms.

\subsubsection{The ``adjusted system"}
\label{sec:adjsystem}

Next, we propose an alternative system
which also tries to control the violation of constraint equations
actively, which we named ``adjusted system".
We think that 
this system is more practical and robust 
than the previous $\lambda$-system.
The essential is summarized as follows: 

\noindent
$\Box $
{\it The Adjusted system (procedures):}  
\begin{enumerate}
\item {Prepare a set of evolution eqs.}
\begin{equation}
\partial_t u=J \partial_i u +K
\end{equation}
\item {Add constraints in RHS }
\begin{equation}
\partial_t u=J \partial_i u +K  \underbrace{+ \kappa C}
\end{equation}
\item 
Choose the coefficient (or Lagrange multiplier)  $\kappa$  so as to make the 
eigenvalues of the homogenized adjusted 
$\partial_t C$ equations negative real value or pure imaginary. 
\begin{eqnarray}
\partial_t C&=&D \partial_iC+E C \\
\partial_t C&=&D \partial_iC+E C \underbrace{+ F \partial_iC + G C} \qquad \Box
\end{eqnarray}
\end{enumerate}
The process of adjusting equations is a common technique in other re-formulating
efforts as we reviewed.  
However, we try to employ the evaluation process of constraint 
amplification factors as an alternative guideline to hyperbolization of the system. 
We will explain these issues in the next section.


\section{A unified treatment: Adjusted System} \label{sec3}
This section is devoted to present our idea of ``asymptotically 
constrained system".  
Original references can be found in 
\cite{ronbun2,adjADM,adjADMsch,adjBSSN}.  

\subsection{Procedures : Constraint propagation equations and Proposals}
\label{secADJ}

Suppose we have a set of dynamical variables   $u^a (x^i,t)$, 
and their evolution equations, 
\begin{equation}
\partial_t u^a = f(u^a, \partial_i u^a, \cdots), \label{ueq}
\end{equation}
and the
(first class) constraints, 
\begin{equation}
C^\alpha (u^a, \partial_i u^a, \cdots) \approx 0.
\end{equation}
Note that we do not require (\ref{ueq}) forms a first order hyperbolic
form. 
We propose  to investigate
the evolution equation of $C^\alpha$ (constraint propagation),
\begin{equation}
\partial_t C^\alpha = g(C^\alpha, \partial_i C^\alpha, \cdots),
 \label{Ceq}
\end{equation}
for predicting the violation behavior of constraints in time evolution. 
We do not mean to integrate 
(\ref{Ceq}) numerically together with the original 
evolution equations (\ref{ueq}), but mean to evaluate them
analytically in advance in order to re-formulate the equations (\ref{ueq}). 

There may be two major analyses of (\ref{Ceq}); 
(a) the hyperbolicity of (\ref{Ceq})
when (\ref{Ceq}) is a first order system, and 
(b) the eigenvalue analysis of the whole RHS in 
(\ref{Ceq}) after a suitable homogenization. 
As we mentioned in \S \ref{sec:hypRemark}, 
one of the problems in the hyperbolic analysis is that it only 
discusses the principal part of the system. 
Thus, we prefer to proceed the road (b). 

\noindent
$\Box${\it Constraint Amplification Factors (CAFs): } \\
We propose  to homogenize (\ref{Ceq}) by a Fourier
transformation,  e.g. 
\begin{eqnarray}
&&\partial_t \hat{C}^\alpha = \hat{g}(\hat{C}^\alpha)
=M^\alpha{}_{\beta} \hat{C}^\beta,
\nonumber \\ 
&&\mbox{where~}
C(x,t)^\rho
=\displaystyle{\int} \hat{C}(k,t)^\rho\exp(ik\cdot x)d^3k,
\label{CeqF}
\end{eqnarray}
then to analyze the set of eigenvalues, say $\Lambda$s,  of the 
coefficient matrix, $M^\alpha{}_{\beta}$, in
(\ref{CeqF}).  We call $\Lambda$s the constraint 
amplification factors (CAFs)
of (\ref{Ceq}). $\qquad \Box$

The CAFs predict the evolution of constraint violations. 
We therefore can discuss the ``distance" to the constraint surface
using the ``norm" or ``compactness" of the constraint violations 
(although we do not have exact definitions of these ``$\cdots$" words). 

The next conjecture seems to be quite useful to predict the 
evolution feature of constraints:

\noindent
$\Box${\it Conjecture on CAFs: } 
\begin{itemize}
\item[(A)] If CAF has 
a {\it negative real-part } (the constraints  are forced to be
diminished), then we see
more stable evolution than a system which has positive CAF.
\item[(B)] If CAF has a {\it non-zero  imaginary-part }
(the constraints are propagating away), then we see
more stable evolution than a system which has 
zero CAF. $\qquad\Box$
\end{itemize}

We found that the system becomes more stable when
more $\Lambda$s satisfy the above criteria. 
(The first observations were in the Maxwell and Ashtekar formulations
\cite{ronbun1,ronbun2}). 

\begin{figure}[t]
\includegraphics[width=75mm]{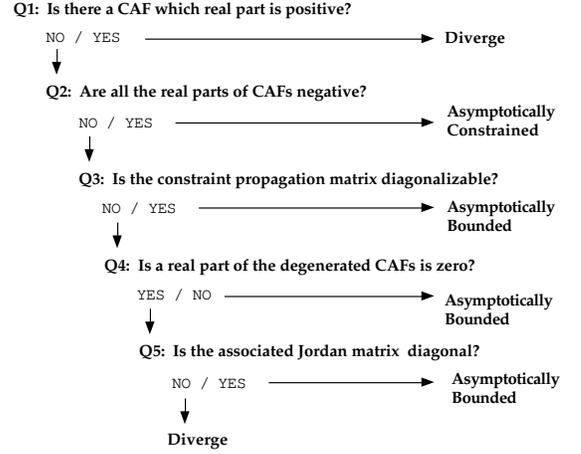}
\caption{A flowchart to classify the constraint propagations. }
\label{fig:flowchart}
\end{figure}

Actually, supporting mathematical proofs are available when we classify
the fate of constraint propagations as follows. \\
\noindent
$\Box${\it Classification of Constraint Propagations: } 
\\
If we assume that avoiding the divergence of constraint norm is related 
to the numerical stability, the next classification would be useful: 
\begin{itemize}
\item[(C1)]  {\it Asymptotically constrained : } 
All the constraints decay and converge to zero. \\ 
This case can be obtained if and only if all the real parts of CAFs are negative. 
\item[(C2)]  {\it Asymptotically bounded : } 
All the constraints are bounded at a certain value. (this includes
the above {\it asymptotically constrained} case.) \\
This case can be obtained if and only if (a) all the real parts of CAFs are not
positive and the constraint propagation matrix $M^\alpha{}_\beta$ is
diagonalizable, or  (b) all the real parts of CAFs are not positive
and the real part of the degenerated CAFs is not zero. 
\item[(C3)]  {\it Diverge : } At least one constraint will diverge. $\qquad \Box$
\end{itemize}
The details are shown in \cite{diagCP}.  
A practical procedure for this
classification is drawn in Figure 
\ref{fig:flowchart}.

The above features of the constraint propagation, (\ref{Ceq}),
will differ when we modify the original evolution equations.
Suppose we add (adjust) the evolution equations using constraints
\begin{equation}
\partial_t u^a = f(u^a, \partial_i u^a, \cdots)
+ F(C^\alpha, \partial_i C^\alpha, \cdots), \label{DeqADJ}
\end{equation}
then (\ref{Ceq}) will also be modified as
\begin{equation}
\partial_t C^\alpha = g(C^\alpha, \partial_i C^\alpha, \cdots)
+ G(C^\alpha, \partial_i C^\alpha, \cdots). \label{CeqADJ}
\end{equation}
Therefore, the problem is how to adjust the evolution equations
so that their constraint propagations satisfy the above criteria 
as much as possible.

\subsection{Applications 1: Adjusted ADM formulations}\label{secADJADM}
\subsubsection{Adjusted ADM equations}
Generally, we can write the adjustment terms to
(\ref{adm_evo1}) and (\ref{adm_evo2})
using (\ref{admCH}) and (\ref{admCM}) by the following combinations
(using up to the first derivatives of constraints for simplicity):

\noindent
$\Box${\it The adjusted ADM formulation \cite{adjADMsch}:}\\
Modify the evolution eqs of $(\gamma_{ij}, K_{ij})$ by constraints 
${\cal H}$ and ${\cal M}_i$, i.e., 
\begin{eqnarray}
{\partial_t} \gamma_{ij} &= (\ref{adm_evo1}) &  +P_{ij} {\cal H}
+Q^k{}_{ij}{\cal M}_k
\nonumber \\&&
+p^k{}_{ij}(\nabla_k {\cal H})
+q^{kl}{}_{ij}(\nabla_k {\cal M}_l), \label{adjADM1}
\\
{\partial_t} K_{ij} &= (\ref{adm_evo2})& +R_{ij} {\cal H}
+S^k{}_{ij}{\cal M}_k
\nonumber \\&&
+r^k{}_{ij} (\nabla_k{\cal H})
+s^{kl}{}_{ij}(\nabla_k {\cal M}_l), \label{adjADM2}
\end{eqnarray}
where $P, Q, R, S$ and $p, q, r, s$ are multipliers.  
According to this adjustment, the constraint propagation equations are 
also modified as
\begin{eqnarray}
\partial_t {\cal H} &=&
\mbox{(\ref{CHproADM})} + \mbox{additional terms}
 \label{Hnewhosei}
\\
\partial_t {\cal M}_i &=&
\mbox{(\ref{CMproADM})} + \mbox{additional terms}
 \qquad \Box \label{Mnewhosei}
\end{eqnarray}


\begin{figure*}
\begin{center}
\includegraphics[width=75mm]{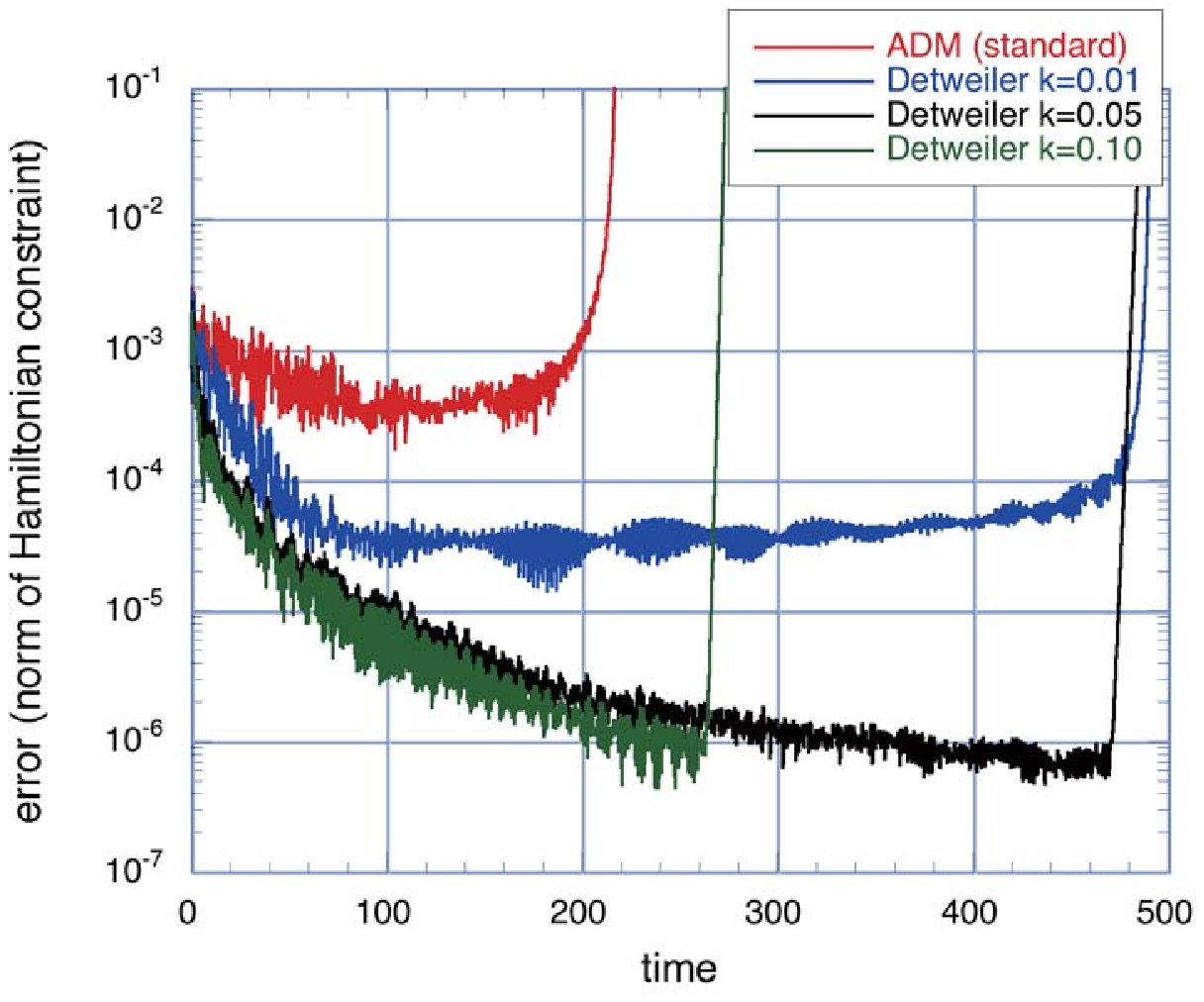} \hspace{5mm}
\includegraphics[width=75mm]{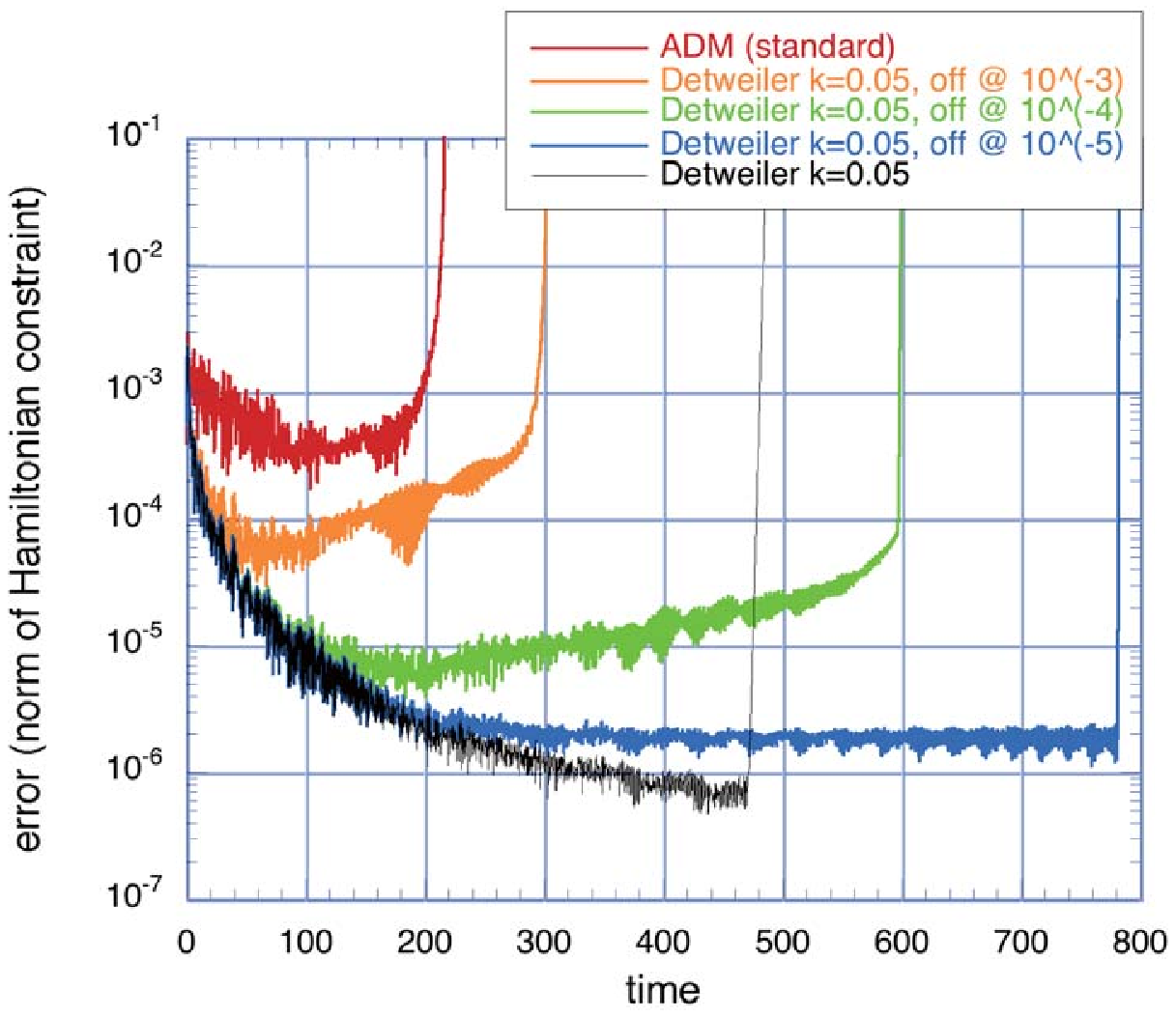}
\caption[quartic]{Demonstration of numerical evolutions
between adjusted ADM systems; especially the standard ADM system and Detweiler's modified ADM system. 
The L2 norm of the constraints ${\cal H}^{ADM}$ 
is plotted in the function of time.  
The model is propagation of Teukolsky wave in the periodical 3-dimensional box. 
$k$ is the parameter in Detweiler's adjustment [$k_L$ in eq.(\ref{Det1})-(\ref{Det4})], 
with fixed-$k$ cases (left panel) and with fixed-and-turnoff-$k$ cases (right panel). 
We see the life-time of simulation becomes four-times longer than the standard ADM
by tuning the parameter $k$. }
\label{figteukwaveADM}
\end{center}
\end{figure*}

We show two examples of adjustments here.  Several others are shown 
in Table 3 of \cite{adjADMsch}.  
\begin{enumerate}
\item {\it The standard ADM vs. original ADM}\\
The first comparison is to show the differences between 
the standard ADM \cite{ADM-York}
and the original ADM system \cite{ADM} (see \S \ref{secADM}).  
In
the notation of (\ref{adjADM1}) and (\ref{adjADM2}), the adjustment, 
\begin{eqnarray}
R_{ij}=\kappa_F \alpha \gamma_{ij}, 
\label{originalADMadjust}
\end{eqnarray}
(and set the other multipliers zero) will distinguish two, 
where $\kappa_F$ is a constant. 
Here $\kappa_F=0$ corresponds to the
standard ADM (no adjustment),
and $\kappa_F=-1/4$ to the original ADM (without any
adjustment to the canonical formulation by ADM).
As one can check by (\ref{Hnewhosei}) and (\ref{Mnewhosei}) 
adding $R_{ij}$ term keeps the constraint propagation in a 
first-order form.  Frittelli \cite{Fri-con} (see also \cite{adjADM}) 
pointed out that
the  hyperbolicity of constraint propagation equations is better 
in the standard ADM system. 
This stability feature is also confirmed numerically, and we set our
CAF conjecture so as to satisfy this difference. 

\item{\it Detweiler type}\\
Detweiler \cite{detweiler} found that with a particular combination, 
 the evolution of the
energy norm of the constraints, ${\cal H}^2+{\cal M}^2$,
can be negative definite
when we apply the maximal slicing condition, $K=0$.
His adjustment can be written in our notation in
(\ref{adjADM1}) and (\ref{adjADM2}), as
\begin{eqnarray}
P_{ij}&=&-  \kappa_L \alpha^3 \gamma_{ij}, \label{Det1}\\
R_{ij}&=& \kappa_L \alpha^3 (K_{ij}- (1 / 3) K \gamma_{ij}), \label{Det2}
\\
S^k{}_{ij}&=& \kappa_L \alpha^2 [
    {3}  (\partial_{(i} \alpha) \delta_{j)}^{k}
- (\partial_l \alpha) \gamma_{ij} \gamma^{kl}], \label{Det3}
\\
s^{kl}{}_{ij}&=&\kappa_L \alpha^3 [
  \delta^k_{(i}\delta^l_{j)}-(1/3) \gamma_{ij}\gamma^{kl}],
  \label{Det4}
\end{eqnarray}
everything else is zero, where $\kappa_L$ is a multiplier.
Detweiler's adjustment, (\ref{Det1})-(\ref{Det4}),
does not put constraint propagation equation
to a first order form, so we cannot discuss
hyperbolicity  or the characteristic speed of the constraints.
From perturbation analysis on Minkowskii
and Schwarzschild space-time, we confirmed 
that Detweiler's system provides better
accuracy than the standard ADM, but only for small positive $\kappa_L$.
\end{enumerate}
We made various predictions how additional adjusted terms will 
change the constraint propagation \cite{adjADM,adjADMsch}. 
In that process, we applied CAF analysis for Schwarzschild spacetime, and searched that  
when and where the negative real or non-zero imaginary eigenvalues of the homogenized constraint propagation matrix appear, and how they depend on the choice of coordinate system and adjustments.
We found that there {\it is} a constraint-violating mode near the horizon
for the standard ADM formulation, and that constraint-violating mode can be 
suppressed by adjusting equations and by choosing an appropriate 
gauge conditions.

\subsubsection{Numerical demonstrations and remarks}

Systematic numerical comparisons are in progress, 
and we show two sample plots here. 
Figure \ref{figteukwaveADM} is the case of Teukolsky wave \cite{Teukolskywave}
propagation under 3-dimensional periodic boundary condition. 
Both the standard ADM system and Detweiler system [one of the adjusted ADM system with
adjustments (\ref{Det1})-(\ref{Det4})] are compared with the same numerical parameters. 
Plots are the 
L2 norm of the Hamiltonian constraint ${\cal H}^{ADM}$,  i.e. the 
violation of constraints, 
and we see the life-time of the standard ADM evolution ends up at $t=200$. 
However, if we chose a particular value of $\kappa_L$ 
[multiplier in (\ref{Det1})-(\ref{Det4})], we
observe that violation of constraints is reduced
than the standard ADM case, and simulation can continue longer than that (left panel).
If we further tuned $\kappa_L$, say turn-off to $\kappa_L=0$ when the total L2 norm of ${\cal H}^{ADM}$
is small, then we can see that the constraint violation is somewhat maintained at a 
small level and more long-term stable simulation is available (right panel).

\begin{figure*}
  \begin{center}
    \begin{tabular}{cc}
      \resizebox{80mm}{!}{\includegraphics{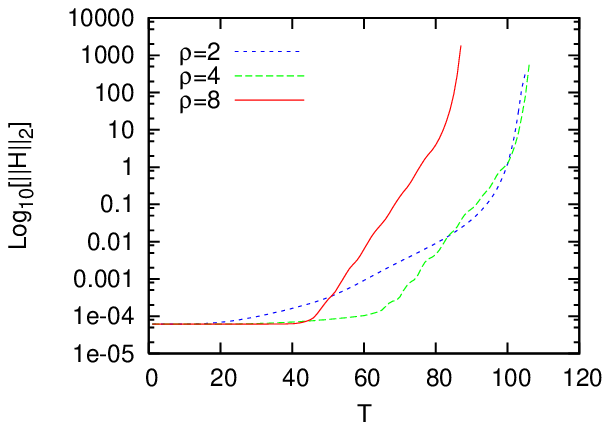}} &
      \resizebox{80mm}{!}{\includegraphics{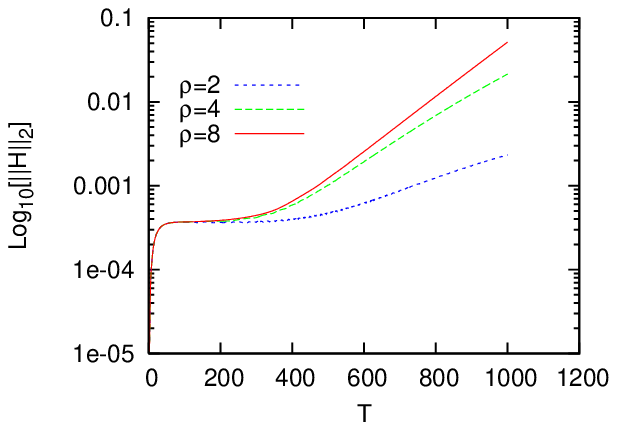}}\\
    \end{tabular}
    \caption{
     The one-dimensional gauge-wave test with the BSSN system (left) and 
     the adjusted BSSN system (right) in 
     the $\tilde{A}$-equation (\ref{B1-adj}). 
     The L2 norm of ${\cal H}$, rescaled by the resolution parameter $\rho^2/4$, are plotted 
     with a function of the crossing-time. The wave amplitude is set to 0.01, and we chose
the adjustment parameter $\kappa_{A}=0.005$. 
     The BSSN system loses convergence at the early time, near the 20 
     crossing-time  
     and it will produce the blow-ups of the calculation in the end, while in the adjusted version
     we see the higher resolution runs show convergence longer, i.e., 
     the 300 crossing-time in ${\cal H}$ and all runs can stably evolve up to the 1000 crossing-time. 
     }\label{GaugewaveBSSN}
  \end{center}
\end{figure*}


During the comparisons of adjustments, we found that it is necessary to 
create time asymmetric structure of evolution equations in order to 
force the evolution on to the constraint surface.  
There are infinite ways of adjusting equations, but we found that if we 
follow the next guideline, then such an adjustment will give us time
asymmetric evolution. 

\noindent
$\Box${\it Trick to obtain asymptotically constrained system}: \\
{\it $=$ Break the time reversal symmetry (TRS) of the evolution equation. }
\begin{enumerate}
\item Evaluate the parity of the evolution equation.\\
By reversing the time ($\partial_t \rightarrow -\partial_t$), there are
variables which change their signatures (parity $(-)$)
 [e.g. $K_{ij}, \partial_t \gamma_{ij}, {\cal M}_i, \cdots$], 
while not (parity $(+)$) [e.g. $g_{ij}, \partial_t K_{ij}, {\cal H}, \cdots$]. 
\item Add adjustments which have different parity of that equation.\\
For example,  for the parity $(-)$ equation $\partial_t \gamma_{ij}$, 
add a parity $(+)$ adjustment $\kappa {\cal H}$. $ \qquad \Box$  
\end{enumerate}

One of our criteria, the negative real CAFs, requires breaking
the time-symmetric features of the original evolution equations. 
Such CAFs are obtained by adjusting the terms which break the 
TRS of the evolution equations, and this is available even at the
standard ADM system. 

\subsection{Applications 2: Adjusted BSSN formulations}\label{secADJBSSN}

\subsubsection{Constraint propagation analysis of the BSSN equations}
\label{secADJBSSNa}
In order to understand the stability property of the BSSN system, 
we studied 
the structure of the evolution equations, 
(\ref{BSSNeqmPHI})-(\ref{BSSNeqmTG}), in detail, especially
how the modifications using the constraints, 
(\ref{BSSNconstraintH})-(\ref{BSSNconstraintS}), affect to the stability
\cite{adjBSSN}. 
We investigated the signature of the eigenvalues of the constraint 
propagation equations, 
and explained that the standard BSSN dynamical equations are balanced 
from the viewpoint of constrained propagations, including a clarification 
of the 
effect of the replacement using the momentum constraint equation, which 
was reported by Alcubierre  {\it et al.}  \cite{potsdam9908}.

Moreover, we  
predicted that several combinations of modifications have a 
constraint-damping nature, and named them {\it adjusted BSSN systems}. 
Several adjusted BSSN systems are proposed in Table II of 
\cite{adjBSSN}.

Yo {\it et al.} \cite{YBS} immediately applied one of our proposals to  
their simulations of stationary rotating black hole, and reported that
one adjustment was contributed to maintain their evolution of 
Kerr black hole ($J/M$ up to $0.9M$) for long time ($t \sim 6000M$).  Their 
results also indicates that the evolved solution is closed to the exact one,
that is, the constrained surface.  

Now, let us make clear some current technical tips listed 
in \S \ref{sec:BSSNproscons} using constraint propagation analysis. 
\begin{itemize}
\item[tip-1]
Trace-out $A_{ij}$ technique can be explained that the 
violation of ${\cal A}$-constraint (\ref{BSSNconstraintA}) 
affects to all other constraint violations.
(See the full set of constraint propagation equations in Appendix of \cite{adjBSSN}.)
\item[tip-2]
Replacement of $\tilde{\Gamma}^i$ enables to maintain 
${\cal G}$-constraint (\ref{BSSNconstraintG}) that delays the
violation of ${\cal H}^{BSSN}$ and ${\cal M}^{BSSN}_i$. 
(Again, the statement comes from the full set of constraint propagation equations. )
\end{itemize}

\subsubsection{Numerical Demonstrations}
\label{secADJBSSNb}
We recently presented our numerical comparisons of
3 kinds of adjusted BSSN formulation\cite{adjBSSNnum}. 
We performed three testbeds: gauge-wave, linear wave, and 
Gowdy-wave tests, proposed by the Mexico workshop \cite{mexico1} on the 
formulation problem of the Einstein equations. 
We observed that the signature of the proposed Lagrange multipliers are
always right and the adjustments improve the convergence
and stability of the simulations.  When the original BSSN system
already shows satisfactory good evolutions (e.g., linear wave test), 
the adjusted versions also coincide with those evolutions; 
while in some cases (e.g., gauge-wave or Gowdy-wave tests) 
the simulations using the adjusted systems last 10 times as long as 
those using the original BSSN equations.

Figure \ref{GaugewaveBSSN} show the comparison between the (plain) BSSN system and 
the adjusted BSSN system in $\tilde{A}$-equation using the 
momentum constraint:  
\begin{eqnarray}
\partial_t \tilde{A}_{ij} = \partial^B_t \tilde{A}_{ij} 
+ \kappa_{A} \alpha \tilde{D}_{(i} {\cal M}_{j)}, \label{B1-adj}
\end{eqnarray}
where $\kappa_{\cal A}$ is predicted (from the eigenvalue analysis)
to be positive in order to damp the constraint violations. 
The testbed is  one-dimensional gauge-wave; the trivial Minkowski space-time but sliced
with the time-dependent 3-metric. 
The poor performance of the plain BSSN system of this test has been already reported
\cite{Jansen:2003uh}, and one remedy is to apply the 4th-order finite 
differencing scheme \cite{Zlochower:2005bj}.
The plots show that our adjusted system also improve the life-time of the plain 
BSSN simulation at least 10 times longer with better convergence.

\section{Outlook}\label{sec:outlook}

\subsection{What we have achieved}
We reviewed recent efforts to the formulation problem of numerical relativity;
the problem to find out a robust system against constraint violations. 
We categorized the approaches into 
\begin{itemize}
\item[(0)]~The standard ADM formulation (\S \ref{secADM}), 
\item[(1)]~The BSSN formulation (\S \ref{secBSSN}), 
\item[(2)]~Hyperbolic formulations (\S \ref{secHYP}), and 
\item[(3)]~Asymptotically constrained formulations (\S \ref{secASYMPT}). 
\end{itemize}
Most of the numerical relativity groups now use the BSSN set of equations which 
are obtained empirically.  A dramatic announcement of the success of 
binary black-hole simulations rushes the community to follow that recipe. 
Actually we are not yet completely understanding why the current set of 
BSSN equations together with particular combinations of gauge condition 
works well.  Several explanations are applied based on the hyperbolic formulation 
scheme, but as we viewed there are not yet satisfactory.

Our approach, on the other hand, tries to construct an evolution system 
that has its constraint surface as an attractor.
Our unified view is to understand the evolution system by evaluating its constraint
propagation.  Especially we proposed to analyze the constraint 
amplification factors  which are the eigenvalues of the homogenized constraint
propagation equations. 
We analyzed the system based on our conjecture  whether the constraint
amplification factors suggest the constraint to decay/propagate or not.  
We concluded that
\begin{itemize}
\item The constraint propagation features become different by simply adding 
constraint terms to the original evolution equations  
(we call this the {\it adjustment} of the evolution equations). 
\item There {\it is} a constraint-violating mode in the standard ADM evolution system
when we apply it to a single non-rotating black hole space-time, and its 
growth rate is larger near the black-hole horizon. 
\item Such a constraint-violating mode can be killed if we adjust the
evolution equations with a particular modification using constraint terms. 
An effective guideline is to adjust terms as they break the time-reversal symmetry
of the equations. 
\item Our expectations are borne out in simple 
numerical experiments using the Maxwell, Ashtekar, and ADM systems.  
However the modifications
are not yet perfect to prevent non-linear growth of the constraint violation. 
\item 
We understand why the BSSN formulation works better than the ADM one
in the limited case (perturbation analysis in the flat background), 
 and further we 
proposed modified evolution equations along the lines of our previous procedure. 
Some of these proposed adjusted systems are numerically confirmed to work better than the
standard BSSN system. 
\end{itemize}

The common key to the problem is how to adjust the evolution equations with 
constraints.  
Any adjusted systems are mathematically equivalent if the constraints are
completely satisfied, but this is not the case for numerical simulations. 
Replacing terms with constraints is one of the normal steps 
when people re-formulate equations in a hyperbolic form. 

In summary, let me answer the following three questions:
\begin{itemize}
\item 
What is the guiding principle for
selecting evolution equations for
simulations in GR?\\
--The key is to analyze the 
constraint propagation equation of the system. 
\item 
Why many groups use the BSSN
equations? \\
--Because people just rush, not to be late to others.
\item 
Are there an alternative
formulation better than the BSSN?\\
--Yes, there are. But we do not know which is the best one yet.
\end{itemize}
\subsection{Future directions}

If we say the final goal of this project is to find a robust evolution system
against violation of constraints, 
then the recipe should be a combination of 
(a) formulations of the evolution equations, 
(b) choice of gauge conditions, 
(c) treatment of boundary conditions, and 
(d) numerical integration methods. 
We are now in the stages of solving these mixed puzzles. 

Recent attentions to higher dimensional space-time studies are waiting for 
numerical researches, but 
it is known that the formulation problem also exists in higher 
dimensional cases \cite{ndimCP}. 

We have written this review from the viewpoint that the
general relativity is a constrained dynamical system.  
This is not a proper problem 
in general relativity, but there is also in many physical systems such as 
electrodynamics, magnetohydrodynamics, 
molecular dynamics, mechanical dynamics. 
Therefore, sharing and interacting the thoughts between different fields will 
definitely accelerate the progress.  

The ideal almighty algorithm to solve all the problems may not exist, 
but the author believe that
our final numerical recipe is somewhat an {\it automatic} system, and 
hope that numerical relativity turns to be an easy {\it toolkit} 
for everyone in near future.

\section*{Acknowledgments}
The author appreciate the LOC of APCTP Winter School on Black Hole Astrophysics 2008, 
for their organization and hospitality.  
The author also thanks 
Gen Yoneda for collaborating our series of works, and Kenta Kiuchi for his recent
numerical experiments. 
The author was partially supported by the Special Research Fund 
(Project No. 4244) of the Osaka Institute of Technology. 
A part of the numerical calculations was carried out 
on the Altix3700 BX2 at YITP at Kyoto University.






\begin{references}


\bibitem{CY9506072} 
A. Abrahams, A. Anderson, Y. Choquet-Bruhat, and 
J. W. York, Jr.,  Phys. Rev. Lett. {\bf 75}, 3377 (1995); 
Class. Quant. Grav. {\bf 14},  A9 (1997). 

\bibitem{Alcubierre}
M. Alcubierre, Phys. Rev. D {\bf 55}, 5981 (1997). 

\bibitem{reviewAlcubierre}
M. Alcubierre, Proceedings of the 17th International Conference on General Relativity and Gravitation (GR17), gr-qc/0412019.


\bibitem{potsdam9908} 
M. Alcubierre, G. Allen, B. Bruegmann, E. Seidel and W-M. Suen, 
Phys. Rev. D {\bf 62}, 124011 (2000). 



\bibitem{potsdam0003} 
M. Alcubierre, B. Bruegmann, T. Dramlitsch, J.A. Font, 
P. Papadopoulos, E. Seidel, N. Stergioulas, 
and R. Takahashi, Phys. Rev. D {\bf  62}, 044034 (2000). 


\bibitem{AM}
M. Alcubierre and J. Mass\'o, Phys. Rev. D {\bf 57}, R4511 (1998).


\bibitem{mexico1}
  M.~Alcubierre {\it et al.} (Mexico Numerical Relativity Workshop 2002 Participants)
  Class.\ Quant.\ Grav.\  {\bf 21}, 589 (2004).

\bibitem{AA2003} 
A. Alekseenko and D. Arnold, Phys. Rev. D {\bf  68}, 064014 (2003).

\bibitem{CY9710041} 
A. Anderson, Y. Choquet-Bruhat, J. W. York, Jr., 
Topol. Methods in Nonlinear Analysis, {\bf 10}, 353 (1997).

\bibitem{AY}
A. Anderson and J. W. York, Jr,  
Phys. Rev. Lett. {\bf 82}, 4384  (1999).


\bibitem{eventhorizon}
P. Anninos, D. Bernstein, S. Brandt, J. Libson, J. Mass\'{o},
E. Seidel, L. Smarr, W-M. Suen, and P. Walker, 
Phys. Rev. Lett. {\bf 74}, 630 (1995). 

\bibitem{ADM} 
R. Arnowitt, S. Deser and C.W. Misner, 
in {\em Gravitation: An Introduction to Current Research}, ed. 
by L.Witten, (Wiley, New York, 1962). 

\bibitem{Ashtekar}
A. Ashtekar, Phys. Rev. Lett. {\bf 57}, 2244 (1986);
Phys. Rev.  {\bf D36}, 1587 (1987).




\bibitem{Goddard}
J. G. Baker, {\it et al.}
Phys. Rev. Lett. {\bf 96}, 111102 (2006); 
Phys. Rev. D {\bf 73}, 104002  (2006).


\bibitem{BS} 
T. W. Baumgarte and S. L. Shapiro, Phys. Rev. D {\bf 59}, 024007 (1999). 

\bibitem{reviewBS} 
T. W. Baumgarte and S.L. Shapiro, Phys. Rept. {\bf 376}, 41 (2003). 

\bibitem{BB} 
J. M. Bardeen, L. T. Buchman, Phys. Rev. D {\bf 65}, 064037 (2002). 

\bibitem{Z4}
C. Bona, T. Ledvinka, C. Palenzuela,
Phys. Rev. D {\bf 66}, 084013 (2002); 
C. Bona, T. Ledvinka, C. Palenzuela, and 
\ifmmode \check{Z}\else \v{Z}\fi{}\'a\ifmmode \check{c}\else \v{c}\fi{}ek, 
{\it ibid.} {\bf 67}, 104005 (2003); 
{\it ibid.} {\bf 69}, 064036 (2004).

\bibitem{BM}
C. Bona, J. Mass\'o, Phys. Rev. D {\bf 40}, 1022 (1989); 
%
Phys. Rev. Lett. {\bf 68}, 1097 (1992).  

\bibitem{BMSS}
C. Bona, J. Mass\'o, E. Seidel and J. Stela, 
Phys. Rev. Lett. {\bf 75}, 600 (1995); 
%
Phys. Rev. D {\bf 56}, 3405 (1997).

\bibitem{cactus1}
C. Bona, J. Mass\'o, E. Seidel, and P. Walker, gr-qc/9804052. 




\bibitem{BFHR} 
O. Brodbeck, S. Frittelli, P. H\"ubner, and O.A. Reula, 
J. Math. Phys. {\bf 40}, 909 (1999). 


\bibitem {reviewBruegmann2008}
B. Bruegmann, Proceedings of GRG18 (at Sydney, Australia, 2007), to be published. 

\bibitem{IBVP-BS} 
L. Buchman and O. Sarbach, Class. Quant. Grav. {\bf 23},   6709 (2006). 




\bibitem{LSU-KST} 
G. Calabrese, J. Pullin, O. Sarbach, and M. Tiglio, 
Phys. Rev. D {\bf 66} 064011 (2002); 
{\it ibid.} {\bf 66}, 041501 (2002). 


\bibitem{UTB}
M. Campanelli, C. O. Lousto, P. Marronetti, Y. Zlochower, Phys. Rev. Lett. {\bf 96}, 111101  (2006); 
M. Campanelli, C. O. Lousto, Y. Zlochower, Phys. Rev. D {\bf 73}, 061501(R) (2006).

\bibitem{Choptuik91}
M. W. Choptuik, Phys. Rev. D {\bf 44}, 3124 (1991). 

\bibitem{criticalbehavior}
M. W. Choptuik,
Phys. Rev. Lett. {\bf 70}, 9 (1993). 


\bibitem{CY9506071} 
Y. Choquet-Bruhat and J. W. York, Jr.,  C. R.  Acad. Sc.
Paris {\bf 321}, S\'erie I, 1089,  (1995).  

\bibitem{detweiler} 
S. Detweiler, Phys. Rev. D  {\bf 35}, 1095 (1987). 

\bibitem{LSU}
P. Diener, {\it et al. }
Phys. Rev. Lett. 96 (2006) 121101.



\bibitem{FrauendienerVogel}
J. Frauendiener and T. Vogel, Class. Quant. Grav. {\bf 22} 1769 (2005).  

\bibitem{FriedrichCFE}
H. Friedrich, 
Proc. Roy. Soc. A {\bf 375}, 169 (1981); 
{\it ibid.} {\bf 378}, 401 (1981). 

\bibitem{FriedrichWeyl}
H. Friedrich, 
Commun. Math. Phys. {\bf 91}, 445 (1983). 


\bibitem{IBVP-FN}
H. Friedrich and G. Nagy, Comm. Math. Phys. {\bf 201}, 619 (1999). 

\bibitem{Fri-con} 
S. Frittelli, Phys. Rev. D {\bf 55}, 5992 (1997). 

\bibitem{FrittelliGomez} 
S. Frittelli and R. Gomez,  J. Math. Phys. {\bf 41}, 5535  (2000). 

\bibitem{FR96}
S. Frittelli  and O. A. Reula,
Phys. Rev. Lett. {\bf 76}, 4667 (1996). 



\bibitem{GGH0707}
D. Garfinkle, C. Gundlach, D. Hilditch, arXiv:0707.0726

\bibitem{Z4lambda} 
C. Gundlach, G. Calabrese, I. Hinder, and J. M. Martin-Garcia,
Class. Quant. Grav. {\bf 22}, 3767 (2005). 

\bibitem{Gundlach0406}
C. Gundlach and J. M. Martin-Garcia,
Phys. Rev. D {\bf 70},  044031 (2004);  {\it ibid.}{\bf 74},  024016 (2006).


\bibitem{HernPHD} 
S. D. Hern, PhD thesis, gr-qc/0004036. 

\bibitem{HeyerSarbach} 
H. Heyer and O. Sarbach, Phys. Rev. D {\bf 70}, 104004  (2004). 









\bibitem{Jansen:2003uh}
  N.~Jansen, B.~Bruegmann and W.~Tichy,
  Phys.\ Rev.\  D {\bf 74}, 084022 (2006).


\bibitem{KST} 
L. E. Kidder, M. A. Scheel, S. A. Teukolsky, 
Phys. Rev. D {\bf 64}, 064017 (2001). 

\bibitem{adjBSSNnum} 
K. Kiuchi and H. Shinkai, 
Phys. Rev. D {\bf 77}, 044010 (2008). 


\bibitem{IBVP-KRSW}
H-O. Kreiss, O. Reula, O. Sarbach and J. Winicour, 
Class. Quant. Grav. {\bf 24}, 5973  (2007).  






\bibitem{LindblomScheel} 
L. Lindblom and M. Scheel, 
Phys. Rev. D {\bf 66},  084014  (2002). 



\bibitem{Miller} 
M. Miller, gr-qc/0008017. 

\bibitem{MTW}
C. W. Misner, K. S. Thorne, J. A. Wheeler, {\em Gravitation}, 
(Freeman, N.Y., 1973). 


\bibitem{NOR}
G. Nagy, O.E. Ortiz and O.A. Reula, Phys. Rev. D {\bf 70}, 044012 (2004).



\bibitem{SN89} 
T. Nakamura and K. Oohara, in {\em Frontiers in Numerical Relativity} 
edited by C. R. Evans, L. S. Finn, and D. W. Hobill 
(Cambridge Univ. Press, Cambridge, England, 1989). 

\bibitem{SN87}
T. Nakamura, K. Oohara and Y. Kojima, Prog. Theor. Phys. Suppl.
{\bf 90}, 1 (1987).




\bibitem{Pretorius} 
F. Pretorius, Phys. Rev. Lett. 95 (2005) 121101; 
Class. Quant. Grav. 23 (2006) S529. 

\bibitem{reviewPretorius}
F. Pretorius, arXiv:0710.1338





\bibitem{IBVP-RRS}
M. Ruiz, O. Rinne and O. Sarbach, 
Class. Quant. Grav. {\bf 24}, 6349  (2007).  

\bibitem{LSU-BSSN}
O. Sarbach, G. Calabrese, J. Pullin, and M. Tiglio, 
Phys. Rev. D {\bf 66}, 064002 (2002). 



\bibitem{SBCSThyper}
M. A. Scheel, T. W. Baumgarte, G. B. Cook, S. L. Shapiro, S. A. Teukolsky,
Phys. Rev. D {\bf 56}, 6320 (1997); 
\bibitem{SBCST98}
M. A. Scheel, T. W. Baumgarte, G. B. Cook, S. L. Shapiro, S. A. Teukolsky,
Phys. Rev. D  {\bf  58}, 044020 (1998).




\bibitem{nakedsingularity}
S. L. Shapiro and S. A. Teukolsky, Phys. Rev. Lett. {\bf 66}, 994 (1991). 

\bibitem{SN}
M. Shibata and T. Nakamura, Phys. Rev. D {\bf 52}, 5428 (1995).


\bibitem{SY-asympAsh} 
H. Shinkai and G. Yoneda, Phys. Rev. D {\bf 60}, 101502 (1999). 

\bibitem{ronbun1} 
H. Shinkai and G. Yoneda, Class. Quant. Grav. {\bf 17}, 4799 (2000). 

\bibitem{adjADMsch} 
H. Shinkai and G. Yoneda, Class. Quant. Grav. {\bf 19}, 1027 (2002). 

\bibitem{novabook}
H. Shinkai and G. Yoneda, 
in {\em Progress in Astronomy and
Astrophysics} (Nova Science Publ).  The manuscript is
available as gr-qc/0209111.

\bibitem{ndimCP}
H. Shinkai and G. Yoneda, Gen. Rel. Grav. {\bf 36}, 1931 (2004).

\bibitem{SiebelHuebner} 
F. Siebel and P. H\"ubner, Phys. Rev. D {\bf 64}, 024021 (2001). 


\bibitem{ADM-SmarrYork} 
L. Smarr, J. W. York, Jr., Phys. Rev. D {\bf 17}, 2529 (1978). 


\bibitem{PennState2006}
C. F. Sopuerta, U. Sperhake, P. Laguna, Class. Quant. Grav. 23 (2006) S579. 

\bibitem{exactsolution}
H. Stephani, D. Kramer, M. MacCallum, C. Hoenselaers, and E. Herlt, {\em Exact Solutions to Einstein's Field Equations} Second ed., (Cambridge, Cambridge Univ. Press, 2003).

\bibitem{Stewart}
J. M. Stewart, Class. Quant. Grav. {\bf 15}, 2865 (1998).





\bibitem{Teukolskywave}
S. A. Teukolsky, Phys. Rev. D {\bf 26}, 745 (1982). 





\bibitem{YBS}
H-J. Yo, T. W. Baumgarte and S. L. Shapiro,  Phys. Rev. D {\bf 66}, 084026 (2002).


\bibitem{ysPRL}
G. Yoneda and H. Shinkai,
Phys. Rev. Lett. {\bf 82}, 263 (1999);
%
Int. J. Mod. Phys. D. {\bf 9}, 13 (2000).

\bibitem{ronbun2} 
G. Yoneda and H. Shinkai, Class. Quant. Grav. {\bf 18}, 441 (2001). 

\bibitem{adjADM} 
G. Yoneda and H. Shinkai, Phys. Rev. D {\bf 63}, 124019  (2001). 

\bibitem{adjBSSN}
G. Yoneda and H. Shinkai, Phys. Rev. D {\bf 66}, 124003  (2002).

\bibitem{diagCP}
G. Yoneda and H. Shinkai, Class. Quant. Grav. {\bf 20}, L31  (2003).


\bibitem{ADM-York} 
J. W. York, Jr., 
in {\em Sources of Gravitational Radiation}, ed. by L. Smarr, 
(Cambridge, 1979). 

\bibitem{Zlochower:2005bj}
  Y.~Zlochower, J.~G.~Baker, M.~Campanelli and C.~O.~Lousto,
  Phys.\ Rev.\  D {\bf 72}, 024021 (2005)

\end{references}
\end{document}